\documentclass[preprint]{elsarticle}

\usepackage{rotating}
\usepackage{float}
\usepackage{lineno,hyperref}

\usepackage{longtable}
\usepackage{multirow}
\journal{Elsevier}
\begin{document}
	
	\begin{frontmatter}
	
\title{Towards the Development of Realistic Botnet Dataset in the Internet of Things for Network Forensic Analytics: Bot-IoT Dataset}

\author[UNSWCanberra]{Nickolaos Koroniotis}
\author[UNSWCanberra]{Nour Moustafa \corref{corrauthor}}
\cortext[corrauthor]{Corresponding author}
\ead{nour.moustafa@adfa.edu.au}
\author[UNSWCanberra]{Elena Sitnikova}
\author[UNSWCanberra]{Benjamin Turnbull}

\address[UNSWCanberra]{University of New South Wales Canberra, Australia}
	
	\begin{abstract}
		The proliferation of IoT systems, has seen them targeted by malicious third parties. To address this, realistic protection and investigation countermeasures need to be developed. Such countermeasures include network intrusion detection and network forensic systems. For that purpose, a well-structured and representative dataset is paramount for training and validating the credibility of the systems. Although there are several network, in most cases, not much information is given about the Botnet scenarios that were used. This paper, proposes a new dataset, Bot-IoT, which incorporates legitimate and simulated IoT network traffic, along with various types of attacks. We also present a realistic testbed environment for addressing the existing dataset drawbacks of capturing complete network information, accurate labeling, as well as recent and complex attack diversity. Finally, we evaluate the reliability of the BoT-IoT dataset using different statistical and machine learning methods for forensics purposes compared with the existing datasets. This work provides the baseline for allowing botnet identificaiton across  IoT-specific networks. The Bot-IoT dataset can be accessed at \cite{2018q}.
	\end{abstract}

	\end{frontmatter}
	\section{Introduction}

	The rapid development of the Internet has been responsible for the emergence of the Internet of Things (IoT), with such prominent examples as smart homes and cities, healthcare systems, and cyber-physical systems. The IoT is a collection of interconnected everyday devices, augmented with light processors and network cards, capable of being managed through web, app or other kinds of interfaces \cite{ronen2017iot}. Naturally, any new technology that is widely adopted by the public, attracts the interest of cyber attackers for exploiting and abusing it using various complex hacking techniques such as Botnets. This has been compounded by the lack of standardization in IoT systems, as well as in the cheap, lightweight and low-powered devices that comprise many of these systems \cite{ronen2017iot}.  One way that IoT networks have been exploited for criminal purposes is in the propagation of Botnet malware, which has been shown to launch DDoS of up to 1.1 TBps \cite{ronen2017iot}\cite{kolias2017ddos}.
	\newline\newline
	With new and unique threats capable of compromising IoT networks and as existing techniques are inadequate to address them, it is important to develop advanced forensics methods for identifying and  investigating adversarial behavior. Network Forensic techniques are widely used for analyzing network traffic and to identify infected devices taking part in major cyber-attacks \cite{pimenta2017cybersecurity}. Additionally, due to the number and nature of IoT devices, the workload of processing collected data would be an ideal application of Big Data analytics. Big Data analytics is a collection of sophisticated analytical processes, which were designed to handle three main problems, variety, velocity and volume \cite{liu2015external}. Since IoT networks generate enormous volumes of data, it is imperative to employ analytical techniques capable of handling them, in a near-real time fashion. As such, the term \textit{‘forensic analytics’} is defined to demonstrate the role of forensics techniques and big data analytics.
	\newline\newline
	Forensic analytics demand big data sources for validating their credibility in IoT networks. In order to develop forensics and intrusion detection solutions that identify and investigate cyber-attacks, the development of a realistic dataset is still an important topic of research \cite{grajeda2017availability}. Over the years, a number of datasets were developed, each of them having its own advantages and disadvantages. Existing datasets, although applicable in some scenarios, introduce various challenges, for example, a lack of reliably labeled data, poor attack diversity such as botnet scenarios, redundancy of traffic, and missing ground truth \cite{2018,sharafaldin2018toward,moustafa2015unsw,2018a}. However, a realistic Botnet traffic dataset in IoT networks has not been effectively designed. The new Bot-IoT dataset addresses the above challenges, by having a realistic testbed, multiple tools being used to carry out several botnet scenarios, and by organizing packet capture files in directories, based on attack types.
	\newline The main contributions of this paper are as follows:
	\begin{itemize}
		\item We design a new realistic Bot-IoT dataset and give a detailed description of designing the testbed configuration and simulated IoT sensors.
		\item We then statically analyze the proposed features of the dataset using Correlation Coefficient and Joint Entropy techniques.
		\item We also evaluate the performance of network forensic methods, based on machine and deep learning algorithms using the botnet-IoT dataset compared with popular datasets.
	\end{itemize}
	This study is structured as follows. The literature review of the study is discussed in Section 2. In section 3, the testbed that was used to develop the Bot-IoT dataset is presented. Sections 4, 5 and 6 discuss the feature extraction process, the benign and malicious scenarios and the statistical and Machine Learning methods used to analyze the dataset respectively. Finally, in section 7, experimental results are presented and discussed, followed by the conclusion of this paper.
	\newline\newline
	\section{Background and Related Work}

	This second outlines and analyses the IoT and forensic techniques-based machine learning that are used in this work. Additionally, information is given about popular network datasets and their limitations for designing forensic techniques in the IoT. 
		\subsection{IoT and forensic analytics-based machine learning}

		 IoT networks are comprised of both typical network elements (including workstation, laptpos and routers) and IoT devices. Machine-to-machine technologies used in conjunction with cloud paradigms, provide the resulting IoT services to users and organizations, in geographically distributed locations \cite{gubbi2013internet}. Popular applications of IoT systems, such as smart cities, smart healthcare and industrial IoT, are complex because they have multiple sensors that need significant processing power and computation to allow individuals to control the IoT elements in large-scale networks \cite{gubbi2013internet}.
		 \newline\newline
		 The vulnerabilities of IoT networks significantly increases with complex cyber-attacks, such as Botnets. Botnets are synchronized networks of compromised machines, called bots \cite{silva2013botnets,khattak2014taxonomy}. In a Botnet, the attacker, called \textit{botmaster}, maintains a bi-directional channel of communication towards the compromised network, through a Command and Control (C\&C) infrastructure. By using the C\&C a botmaster issues commands to the rest of the botnet and is capable of receiving feedback regarding attacks in progress or even stolen data \cite{silva2013botnets,khattak2014taxonomy}. A botnet can launch a number of activities, such as Distributed Denial of Service attacks (DDoS), Keylogging, Phishing, Spamming, Click fraud, Identity theft and even the proliferation of other Bot malware \cite{amini2015survey}. The compromised bots are under the complete control of the botmaster. 
		 \newline\newline
	     Digital Forensics is defined as “the Identification, Preservation, Collection, Examination, Analysis and Presentation of digital evidence” \cite{palmer2001road}. During this process, forensic methods are applied to data (or in this case, big data), in order to draw conclusions about the occurrence of a crime, including who did it, their motives, what methods they used and what, if any, information or machines they affected. Since the forensic process is linked with analyzing big data to design effective techniques in law enforcement, such techniques can be utilized using Machine Learning (ML) algorithms for designing reliable models that can efficiently determine cybercrime. Therefore, we entitle forensic techniques that use Machine Learning and big data as ‘forensic analytics’. Forensic analytics could be used in the examination phase, where the Forensic analyst seeks to identify patterns that would answer the aforementioned questions, related to the occurrence of a crime.
		\newline\newline
		Network Forensics analytics is tasked with processing network data, such as logs, e-mails and network packets, the latter usually stored in the form of packet capture files.e. In literature, the common cyber applications of network forensics -based ML are Intrusion detection, Honeypots, network flow analysis, deep packet inspection, and email authorship, as discussed in the following points:
		\begin{itemize}
			\item \textbf{Intrusion Detection Systems (IDS)} - \cite{hodo2016threat,garcia2009anomaly,wang2016attack} are classified into network and host intrusion detection systems. Network IDS (NIDS) are placed in a strategic point of the network, usually a node connecting the internal network to the Internet and are tasked with scanning inbound and outbound traffic for known patterns of attacks. As known malicious patterns are identified, notifications are set off and subsequent forensic mechanisms can be deployed, thus lowering response time and potentially increasing accuracy of investigation. Such IDS can incorporate ML, in their anomaly detection process \cite{garcia2009anomaly}, with classification, clustering and other methods used to detect unusual non-normal traffic. Examples of such systems have already been proposed \cite{wang2016attack}.
			\item  \textbf{Honeypots}- are entities (software and/or hardware) that are designed to appear as an appealing, exploitable target to an attacker \cite{wang2016attack,rieck2008learning}. Honeypots have several uses, to identify attackers that are able to bypass other  control mechanisms,to mitigate ongoing attacks and distract attackers from real systems and collect attack binaries \cite{rieck2008learning}. They are classified in three main categories, (i) Low Interaction Honeypots, where software simulates responses to certain stimuli, (ii) High Interaction Honeypots, where real Operating Systems are used as the honeypot, and (iii) Hybrid Honeypots a combination of (i) and (ii). In the context of honeypots, Machine Learning techniques are applied either to identify network traffic patterns that can be used to identify similar attacks in the future (by recording pcaps, collecting logs,…), or to analyse Malware samples\cite{rieck2008learning} obtained during an infection.
			\item \textbf{Network Flow Analysis and Deep Packet Inspection} - can be categorized in two broad groups: (i) Deep Packet Inspection and (ii) Network Flow Analysis \cite{doshi2018machine,nguyen2008survey,pimenta2017cybersecurity}. Both categories use captured traffic to draw conclusions regarding cyber-attacks. Their main difference is that (i) investigates the content of each packet, which yields more complete results but adds a considerable overhead whereas (ii) aggregates packets into flows, which are groups of packets that share {Source IP, Source Port, Destination IP, Destination Port, Protocol}. In both categories, Clustering, Classification and Association Rule Mining techniques have been applied in literature.
			\item \textbf{Email Authorship}- is an example of ML application in the context of Cyber Forensics, not directly related to Network Forensics. In such cases, Support Vector Machines have been used to identify the author of an e-mail based on the text itself \cite{de2001mining}.
		\end{itemize}
		
		\subsection{Comparison of proposed testbed with others}

		In this section we present the comparison between known testbeds found in literature, to the one we used to produce IoT-Bot dataset. A testbed is as an environment designed for the generation of network traffic \cite{moustafa2015unsw}.
		\newline\newline
		A number of datasets have been introduced in literature, to assist researchers in simulating Botnet activities and generating attack traffic datasets \cite{sharafaldin2018toward, moustafa2015unsw,doshi2018machine, alomari2014design,carl2006using,bhatia2014framework,behal2017detection}. To produce these datasets, various different testbeds were used. To start with, similar to our approach, Alomari et al. \cite{alomari2014design} relied on a high-tier Server, outfitted with Virtual Machines to implement their HTTP DDoS Botnet traffic testbed. Although, their testbed employed a larger quantity of Bots (40) than ours, but they limited their generated Botnet activities to only HTTP DDoS attacks, whereas we did not (description of activities in Section 5). The stated goal of this network testbed and associated dataset was to produce traces of malicious traffic. Whilst this is important, the lack of full network packet capture makes it impossible to provide verification of any outcomes. Network traces also limits the data that can be extracted and processed.
		\newline\newline
		Choosing to focus on the C2 (Command and Control) activities of a Botnet, Carl et al. \cite{carl2006using}, implemented their own IRC-centric testbed, in order to produce a ML approach to effectively identify malicious IRC traffic. In their work, they made use of a re-implementation of a real-world Botnet named “Kaiten”, which used IRC traffic as its C\&C communication medium, and launched UDP DDoS from 13 “zombies” targeting a victim host. Contrary to their approach, in our testbed, we mixed normal and attack traffic by activating Ostinato \cite{emaN} (for normal traffic) and the attack software at the same time, whereas their testbed was not tasked with generating normal traffic, with it being collected from a public network (later anonymized) and Botnet traffic being generated by their testbed. A good dataset should include both attack and normal traffic.
		\newline\newline
		Taking a different approach to building their testbed, Bhatia et al. \cite{bhatia2014framework} employed physical machines arranged appropriately in a local network. Their testbed was tasked with simulating flash events and various types of DDoS attacks, the latter through the use of specialized software called Botloader and IP-Aliasing. Compared to our virtualized approach, their choice of using physical machines incurs added costs (for the physical machines), is not easily deployable as mentioned by the research team itself \cite{bhatia2014framework} and does not include the added resiliency that virtualized environments offer. Additionally, our approach included a greater variety of Botnet activities, including but not limited to DDoS, DoS and Port Scanning.
		\newline\newline
		Similarly to \cite{bhatia2014framework}, Behal et al. \cite{behal2017detection}, developed their testbed, DDoSTB to generate DDoS attacks and flash events. Their approach was to use a combination of real computers arranged locally into subnetworks, and emulation software that created multiple virtual nodes per physical. As mentioned for \cite{bhatia2014framework}, the use of physical machines in such a testbed lacks the resiliency, ease and speed with which Virtual Machines can be deployed. Also, again the focus of the testbed was DDoS attacks, which, although a significant and particularly destructive capability of Botnets is far from the only kind of action that they can perform, as has been observed, that Botnets in the wild exhibit a number of diverse malicious actions, such as Keylogging, Phishing, Data theft, DDoS, DoS and Malware proliferation.
		\newline\newline 
		Comparable to \cite{bhatia2014framework,behal2017detection} but more elaborate in their implementation, Sharafaldin et al. \cite{ sharafaldin2018toward} crafted a testbed relying on physical machines, which were separated into two networks, the Victim-network and the Attack-network. Their approach made use of several popular Operating Systems, such as Windows (7Pro, 8.1, Vista), Ubuntu (16.4,14.4, Server) and Kali Linux, a choice that mirrors our own. Moustafa et al. \cite{moustafa2015unsw} relied on the IXIA Perfect Storm in their testbed implementation to generate both normal and malicious network traffic for the purpose of generating the UNSW-NB15 dataset. Through IXIA, the researchers installed three virtual servers, two of which were tasked with generating normal traffic with the third performing attacks on them.
		\newline\newline
		Doshi et al. \cite{doshi2018machine} work focused on generating a ML model that would identify IoT-produced DDoS attacks. In their approach, they made use of physical IoT consumer appliances, which were deployed in a local network. Normal traffic was collected by interacting with the IoT devices, while DoS attacks were simulated. The main differences of their proposed testbed and the one we implemented, is scale and attack variety. For our testbed, we chose to use four Kali Linux machines to simulate both DoS and DDoS attacks, along with other Botnet actions.
		\newline\newline
		The main novelty of the proposed dataset, is the introduction of the IoT element in the environment. Namely, we employed popular middleware (Node-red), to simulate the existence of IoT devices in our Virtual Network. This IoT simulated traffic is in the form of MQTT, a publish-subscribe communication protocol implemented over TCP/IP, often used for lightweight network communication, such as IoT \cite{soni2017survey}. In contrast, IoT devices were not represented in the testbeds that were presented in the previous section \cite{sharafaldin2018toward,moustafa2015unsw}\cite{alomari2014design,carl2006using,bhatia2014framework, behal2017detection} with the exception of \cite{doshi2018machine}.
		\newline\newline
		Choosing this virtualized setup, carries a number of pros and cons. To start with the pros, using virtualized environment means that the setup is portable and easy to set up with relatively low cost. Additionally, using simulations to represent the servers, PCs and IoT devices made the launching of malicious attacks easier and safer, with the extra bonus that the machines could be recovered easily if necessary. From a security perspective, not using actual Botnet malware to infect our machines made the process a lot safer, as by doing so we ran the risk of either unwillingly taking part in real attacks (our machines become part of a real botnet).\newline\newline
		Furthermore, many newer versions of Bot malware can detect a virtualized environment, producing fake data as subterfuge. With regards to the generation of realistic normal network traffic, the Ostinato \cite{emaN} Software was used, to make the dataset appear as if it were collected from a real-world network. Lastly, we executed a set of standard Botnet attacks, which makes this dataset useful for the development of realistic Botnet traffic detection.
		\newline\newline
		On the other hand, using virtualized environment prevents us from launching in-depth attacks against an IoT’s firmware and hardware, thus somewhat limiting the depicted ways through which an attack against such machines can be launched. Nevertheless, the collected dataset is adequate for our purposes. In addition, as an expansion, an even more diverse group of attacks could be performed, such as application layer DoS/DDoS, Layer 1,2 attacks on physical IoT devices, something which requires the use of real IoT devices, access to which we did not have at the time of the experiments.
		
		\subsection{Existing network datasets and their forensics analytics limitations}
		
		\begin{center}
			\begin{small}
				\begin{longtable}{|p{0.1\textwidth}|p{0.1\textwidth}|p{0.1\textwidth}|p{0.1\textwidth}|p{0.1\textwidth}|p{0.1\textwidth}|p{0.1\textwidth}|p{0.1\textwidth}|}
					\caption{\normalsize Comparison of datasets 
						(T=true, F=false)}
					\label{table:ComparisonOfDatasets}\\
					
					\hline Dataset&Realistic testbed configuration&Realistic traffic&Labeled data&IoT traces&Diverse attack scenarios&Full packet capture&New generated features\\ \hline
					
					Darpa98&T&F&T&F&T&T&F\\ \hline				
					KDD99&T&F&T&F&T&T&T\\ \hline					
					DEFCON-8&F&F&F&F&T&T&F\\ \hline					
					UNIBS&T&T&T&F&F&T&F\\ \hline					
					CAIDA&T&T&F&F&F&F&F\\ \hline
					LBNL&F&T&F&F&T&F&F\\ \hline
					UNSW-NB15&T&T&T&F&T&T&T\\ \hline
					ISCX&T&T&T&F&T&T&T\\ \hline
					CICIDS 2017&T&T&T&F&T&T&T\\ \hline
					TUIDS&T&T&T&F&T&T&T\\ \hline
					Bot-IoT&T&T&T&T&T&T&T\\ \hline
					
				\end{longtable}
			\end{small}
		\end{center}
		Since the applications of forensics discussed in the IoT and Forensic analytics section employ machine learning techniques, they require big datasets for analyzing network flows, differentiating between normal and abnormal traffic, and producing forensic reports, which could be useful to forensic specialists in law enforcement. The development of a realistic network dataset is a very important task for designing network forensics, intrusion detection and privacy-preserving models. Over the years, several datasets have been produced \cite{sharafaldin2018toward} and although a good number of them remain private, due to primarily privacy concerns, some have become publicly available. The most commonly used datasets are briefly explained below, with a comparison between them and Bot-IoT given in Table \ref{table:ComparisonOfDatasets}.
		\begin{itemize}
			\item The DARPA 98 dataset was generated by MIT’S Lincoln Lab for assessing intrusion detection systems. The resulting dataset was produced in a period of 7 weeks, was made up of 4GB of binary data and simulated a small Air Force network connected to the Internet \cite{brugger2007assessment}\cite{fernandezugr}, which was later enhanced in 1999 to generate the features in the KDD99 dataset.
			\item The KDD99 dataset was generated from the DARPA 98 dataset for evaluating intrusion detection systems that distinguish between inbound attacks and normal connections \cite{2018}\cite{tavallaee2009detailed}\cite{fernandezugr}. Even though it is still used to this day, it has several problems, for example, non-normal distributions of attack and normal data named the imbalanced learning problem. The NSL-KDD dataset was proposed to address the limitations of the KDD99, but the two versions are outdated and do not reflect current normal and attack events \cite{tavallaee2009detailed}\cite{sharafaldin2018toward}.
			\item The DEFCON-8 dataset consists of port scanning and buffer overflow attacks, which were recorded during a “Capture the Flag” competition \cite{sharafaldin2018toward}. As it lacks a significant quantity of normal background traffic, its applicability for evaluating Network Forensics and IDS systems is limited.
			\item The UNIBS \cite{ UNIBS2009}\cite{bhuyan2015towards} dataset was developed by the University of Brescia, Italy. In their configuration, the researchers installed 20 workstations running the Ground Truth daemon and traffic was collected through tcpdump at the router to which they were collected. Although the researchers used a realistic configuration, there are some drawbacks to this dataset. First, the attack scenarios are limited to DoS attacks. Secondly, the dataset exists in packet form with no extra features generated on it. Additionally, no information is given about the labels.
			\item The CAIDA datasets \cite{bhuyan2015towards}\cite{2018e} are collections of varied data types. They are comprised of anonymized header traffic, excluding the payload. The datasets are made up of very specific attacks, such as DDoS. One popular dataset from the CAID collection, is the CAIDA DDoS 2007, which includes one hour of anonymized attack traces from DDoS attacks that took place in August 4 2007. One drawback of the CAIDA datasets is that they did not have a ground truth about the attack instances. Additionally, the gathered data was not processed to generate new, features which could improve the differentiation between attack and normal traffic.
			
			\item The LBNL \cite{2005}\cite{bhuyan2015towards} dataset consists of anonymized traffic, which consists of only header data. It was developed at the Lawrence Berkley National Laboratory, by collecting real inbound, outbound and routing traffic from two edge routers. Similarly to the UNIBS, the labeling process is lacking and no extra features were generated, with the data existing as a collection of .pcap files. 
			\item The UNSW-NB15 is a dataset developed at UNSW Canberra by Mustafa et al. \cite{moustafa2015unsw}. The researchers employed IXIA perfect storm to generate a mixture of benign and attack traffic, resulting in a 100GB dataset in the form of PCAP files with a significant number of novel features generated. The purpose of the produced dataset was to be used for the generation and validation of intrusion detection. However, the dataset was designed based on a synthetic environment for generating attack activities.
			\item The ISCX dataset \cite{CanadianInstituteofCybersecurity2018}\cite{ammar2015decision} was produced at the Canadian Institute for Cyber security. The concept of profiles was used to define attack and distribution techniques in a network environment. Several real traces were analyzed to generate accurate profiles of attacks and other activities to evaluate intrusion detection systems. Recently, a new dataset was generated at the same institution, the CICDS2017 \cite{sharafaldin2018toward}. The CICIDS2017 is comprised of a variety of attack scenarios, with realistic user-related background traffic generated by using the B-Profile system. Nevertheless, the ground truth of the datasets, which would enhance the reliability of the labeling process, was not published. Furthermore, applying the concept of profiling, which was used to generate these datasets, in real networks could be problematic due to their innate complexity. 
			\item The TUIDS \cite{gogoi2012packet}\cite{bhuyan2015towards} dataset was generated by the Tezpur University, India. This dataset features DDoS, DoS and Scan/Probing attack scenarios, carried out in a physical testbed. However, the flow level data do not include any new features other than the ones generated by the flow-capturing process.
		\end{itemize}
		Although various research studies have been conducted \cite{2018,sharafaldin2018toward,moustafa2015unsw,UNIBS2009,2005,gogoi2012packet,2018e,CanadianInstituteofCybersecurity2018} to generate network datasets, the development of realistic IoT and network traffic dataset that includes recent Botnet scenarios still is an unexplored topic. More importantly, some datasets lack the inclusion of IoT-generated traffic, while others neglected to generate any new features. In some cases, the testbed used was not realistic while in other cases, the attack scenarios were not diverse. This work seeks to address the shortcomings by designing the new Bot-IoT dataset and evaluate it using multiple forensics mechanisms, based on machine and deep learning algorithms.

	\section{The proposed IoT-Bot dataset}
		\subsection{Overview of proposed testbed}

		The proposed testbed consists of three components, namely: network platforms, simulated IoT services, and extracting features and Forensics analytics. First, the network platforms include normal and attacking virtual machines (VMs) with additional network devices such as a firewall and tap. Second, the simulated IoT services, which contain some IoT services such as a weather station. These are simulated through the Node-red tool \cite{2018i}. Third, extracting features and forensics analytics, where the Argus tool \cite{2018j} was used in order to extract data features, and afterward statistical models and machine Learning techniques were employed in order to assess the feature vectors for discriminating normal and abnormal instances. More details of the components are discussed below.
		\begin{figure}[H]
			\centering
			\includegraphics[width=\textwidth]{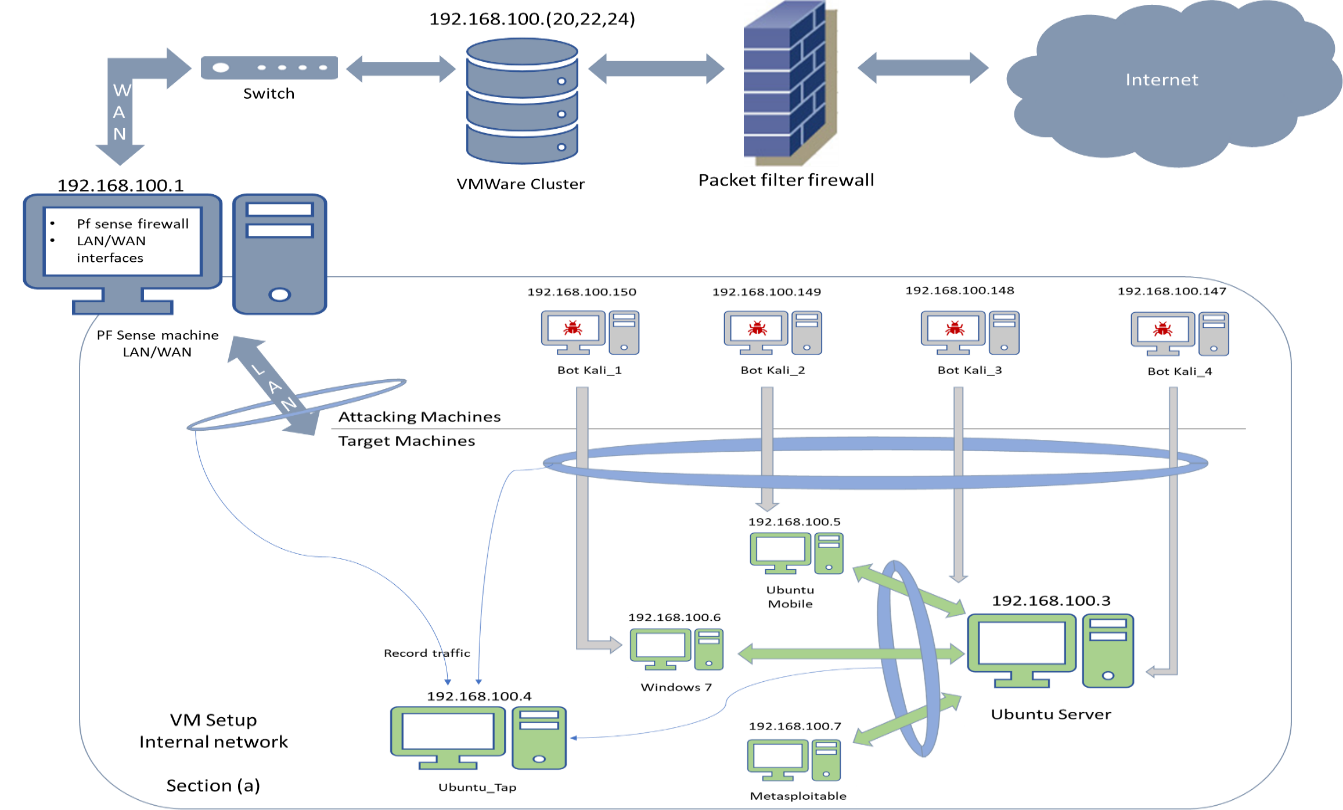}
			\caption{Testbed environment of the new Bot-IoT dataset}
		\end{figure}
		\subsection{Network platforms}

		We designed the testbed environment at the Research Cyber Range lab of UNSW Canberra. Virtual Machines that were prepared for this task, were ported into an ESXi \cite{2018k} configured cluster and managed through a vSphere platform \cite{2018l}. In Figure 1, we depict the testbed of the Bot-IoT dataset, where several VMs are connected to LAN and WAN interfaces in the cluster and are linked to the Internet through the PFSense machine. On the Ubuntu VM platforms, the Node-red tool was used for simulating various IoT sensors which were connected with the public IoT hub, AWS \cite{2018m}. We developed Java scripts on the Node-red tool for subscribing and publishing IoT services to the IoT gateway of the AWS via the Message Queuing Telemetry Transport (MQTT) protocol \cite{soni2017survey}, as detailed in the following sub-section.
		\newline\newline
		There were also a packet filtering firewall and two Network Interface Cards (NICs) configured in the environment. One of the NIC was configured for LAN and the other one for WAN. The main reason for using this firewall is to ensure the validity of the dataset labeling process, as it enables to manage network access by monitoring incoming and outgoing network packets, based on specific source and destination internet protocol (IP) addresses of the attack and normal platforms. VMs which needed to communicate with the Internet, sent their traffic through the PFSense machine which in turn, forwarded the traffic through a switch and a second firewall before it could be routed further into the Internet.
		\newline\newline
		Our network of VMs consists of four Kali machines, Bot Kali\textunderscore1, Bot Kali\textunderscore2, Bot Kali\textunderscore3, and Bot Kali\textunderscore4, an Ubuntu Server, Ubuntu mobile, Windows 7, Metasploitable and an Ubuntu Tap machine. The Kali VMs, which belong to the attacking machines, performed port scanning, DDoS and other Botnet-related attacks by targeting the Ubuntu Server, Ubuntu mobile, Windows 7 and Metasploitable VMs. In the Ubuntu Server, a number of services had been deployed, such as DNS, email, FTP, HTTP, and SSH servers, along with simulated IoT services, in order to mimic real network systems.
		\newline\newline
		To generate a massive amount of normal traffic, we used the Ostinato tool \cite{emaN}, due to its flexibility of generating realistic benign traffic with given IPs and ports. We also maintained periodically normal connections between the VMs by executing normal functions of the services installed on the Ubuntu server, such examples include the DNS server, which resolved the names of the VMS to their IPs and the FTP server, used to transfer particular files between the VMs. To collect the entire normal and attack raw packet volume exchanged within the configured network, the tshark tool was used on the Ubuntu Tap machine, by setting its NIC in a promiscuous mode that ensured the scalability of the testbed. 
		
		\subsection{Simulated IoT services}

		In order to simulate the network behavior of IoT devices, we employed the Node-red tool \cite{2018i}. Node-red is a popular middleware used to connect IoT physical devices with their backend cloud server and applications, improving and speeding up communications between the various parts of an IoT deployment. On the Node-Red tool, we developed JavaScript code that mimicked IoT sensors such as temperature, pressure and humidity sensors.
		\begin{figure}[H]
			\centering
			\includegraphics[width=\textwidth]{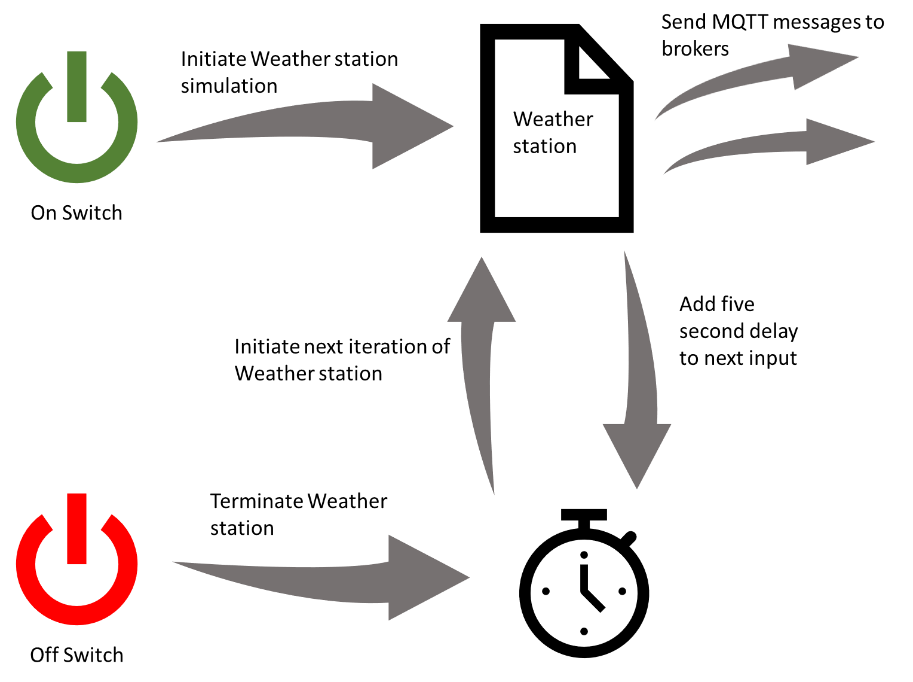}
			\caption{Flowchart of weather IoT simulation of the node-red tool used in the dataset}
		\end{figure}

	The various pieces of code were triggered for publishing and subscribing to the topics, as shown in the example of Figure 2. The MQTT protocol \cite{soni2017survey} was used as a lightweight communication protocol that links machine-to-machine (M2M) communications, making it a viable choice for IoT solutions. It works in a publish/subscribe model, where a device publishes data to an MQTT broker (server side) under a topic, which is used to organize the published data and allows clients to connect to the broker and fetch information from the topic of the device they wish to interact with. 
	\newline
	We applied the following IoT scenarios in the testbed of the dataset:
	\begin{enumerate}
		\item A weather station (Topic:\textit{/smarthome/weatherStation}), which generates information on air pressure, humidity and temperature.
		\item A smart fridge (\textit{/smarthome/fridge}), which measures the fridge’s temperature and when necessary adjusts it below a threshold.
		\item Motion activated lights (\textit{/smarthome/motionLights}), which turn on or off based on a pseudo-random generated signal. 
		\item A remotely activated garage door (\textit{/smarthome/garageDoor}), which opens or closes, based on a probabilistic input.
		\item A smart thermostat (\textit{/smarthome/thermostat}), which regulates the house’s temperature by starting the Air-conditioning system.
	\end{enumerate}
	The five IoT scenarios were connected to both the Ubuntu server, where the Mosquitto MQTT broker \cite{2018n} was installed, as-well-as the AWS IoT hub. While running the testbed environment, MQTT messages were published periodically from all clients to both brokers. The connections allowed us to simulate regular IoT traffic, since the MQTT brokers were used as intermediaries that connected smart devices with web/smartphone applications.
	\section{Extracting features and forensics analytics}

	After collecting the pcap files from the virtualized setup, with normal and attack traffic in the same files, we extracted the flow data, by using the Argus tools and produced .argus files. Following the flow extraction process, the produced flow data was imported into a MySQL database for further processing. We then employed statistical measures using Correlation Coefficient \cite{hall2015pearson} and Entropy \cite{lesne2014shannon} techniques to assess the original dataset and select important feature, as described in Section 6. New features were generated based on the transactional flows of network connections in order to discover normal and intrusive events. Finally, three Machine Learning models, which could be applied for forensic purposes, were trained and validated on several versions of the dataset to assess the value of the dataset compared with other datasets, as discussed in section 7.
		\subsection{Network Flow extraction process}
		Capturing network traffic while ensuring the labeling process is not an easy task, as the synchronization of sending and receiving packets and later tagging these packets either normal or attack should be timely and automatically developed. In order to accomplish this task, we developed some scripts on the Cron Linux functions \cite{2018o} over the Ubuntu Tap VM. When the scripts ran on a given time, a particular normal or attack scenario had to be executed. For example, during the generation of DDoS, we scheduled the execution of custom bash scripts which invoked hping3 and golden-eye to run the DDoS attacks, while simultaneously normal traffic was generated in the background. At the same time, the tshark tool \cite{2018p} was running, to capture raw packets and store them in 1 GByte pcap files to ease extracting network features.
		\newline\newline
		We scheduled different types of attacks to run at different times, with normal background traffic being constantly generated. By doing so, we ensured that different types of attacks would be performed at different times allowing us to organize the produced pcap files, based on attack category and subcategory. For this purpose, the attacking Kali Bots 1-4 and the recording Ubuntu Tap, had to be synchronized, so that Ubuntu Tap could halt the recording of a particular attack’s pcap files and start the next one scheduled. The normal traffic, which was mixed with the attack traffic was generated by the Ostinato \cite{emaN} program that ran in the Ubuntu\textunderscore Server VM. Knowing the IP addresses of both attacker and victim machines, enabled us to differentiate between normal and attack traffic, as we ensured that between the two groups, only attacking traffic would be transferred.
		
		\begin{figure}[H]
			\centering
			\includegraphics[width=\textwidth]{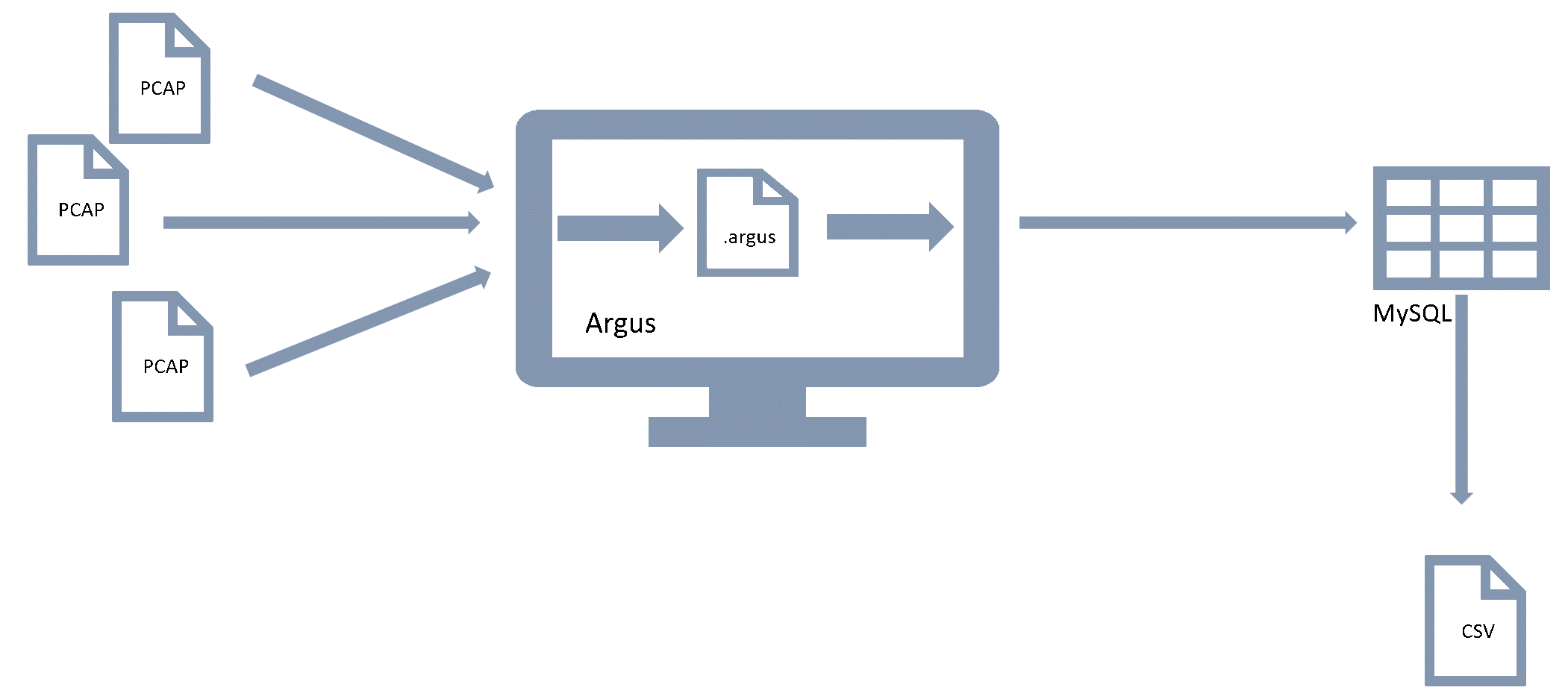}
			\caption{Process of converting pcap to final csv dataset}
		\end{figure}
		After collecting the pcap files, the Argus tool \cite{2018b} was used to generate the relevant network flows. The process can be viewed in Figure 4. The pcap files were converted into Argus format by using the Argus client program. Then, the \textit{rasqlinsert} command was applied to extract network flow information, and simultaneously log the extracted features into MySQL tables. \newline
		The final features produced by Argus during the Network flow extraction process are listed in Table ~\ref{table:FeaturesAndDescriptions}.
		
		\begin{center}
		\begin{small}
			\begin{longtable}{|p{0.15\textwidth}|p{0.34\textwidth}|}
				\caption{\normalsize Features and descriptions}
				\label{table:FeaturesAndDescriptions}\\
				
				\hline Feature&Description\\ \hline
				pkSeqID&Row Identifier\\ \hline
				Stime&Record start time\\ \hline
				flgs&Flow state flags seen in transactions\\ \hline
				flgs\textunderscore number&Numerical representation of feature flags\\ \hline
				Proto&Textual representation of transaction protocols present in network flow\\ \hline
				proto\textunderscore number&Numerical representation of feature proto\\ \hline
				saddr&Source IP address\\ \hline
				sport&Source port number\\ \hline
				daddr&Destination IP address\\ \hline
				dport&Destination port number\\ \hline
				pkts&Total count of packets in transaction\\ \hline
				bytes&Totan number of bytes in transaction\\ \hline
				state&Transaction state\\ \hline
				state\textunderscore number&Numerical representation of feature state\\ \hline
				ltime&Record last time\\ \hline
				seq&Argus sequence number\\ \hline
				dur&Record total duration\\ \hline
				mean&Average duration of aggregated records\\ \hline
				stddev&Standard deviation of aggregated records\\ \hline
				sum&Total duration of aggregated records\\ \hline
				min&Minimum duration of aggregated records\\ \hline
				max&Maximum duration of aggregated records\\ \hline
				spkts&Source-to-destination packet count\\ \hline
				dpkts&Destination-to-source packet count\\ \hline
				sbytes&Source-to-destination byte count\\ \hline
				dbytes&Destination-to-source byte count\\ \hline
				rate&Total packets per second in transaction\\ \hline
				srate&Source-to-destination packets per second\\ \hline
				drate&Destination-to-source packets per second\\ \hline
				attack&Class label: 0 for Normal traffic, 1 for Attack Traffic\\ \hline
				category&Traffic category\\ \hline
				subcategory&Traffic subcategory\\ \hline
				
				\end{longtable}
			\end{small}
		\end{center}
		Additionally, the group \lq saddr\rq, \lq sport\rq, \lq daddr\rq, \lq dport\rq, \lq proto\rq are considered network flow identifiers, as this information is capable of uniquely identifying a flow at any given time and assisting in the labeling process. To label the data for use with machine learning processes, we employed \lq alter table\rq queries to introduce the new columns and \lq update\rq queries to modify the values based on saddr and daddr values. In the dataset, attack instances are labeled with \lq 1\rq and normal ones are labeled with \lq 0\rq for training and validating machine learning models through a binary classification. In addition to that, we have further introduced attack category and subcategory attributes, which could be used for training and validating multiclass classification models.

		\subsection{New feature generation}
		
		We developed new features that were generated based on the features listed in Table \ref{table:FeaturesAndDescriptions}. The main purpose of this process is to improve the predictive capabilities of classifiers. The new features demonstrated in Table ~\ref{table:GeneratedFlowFeatures} were designed over a sliding window of 100 connections. The number 100 although chosen arbitrarily, pays a significant role in the generation of these new features, as it captures the statistics of groups of flows, in a relatively small time-window, inside of which, patterns of several attacks can be discovered. In order to generate these features in MySQL DB, we made use of stored procedures.
		
		\begin{center}
			\begin{small}
				\begin{longtable}{|p{0.02\textwidth}|p{0.4\textwidth}|p{0.34\textwidth}|}
					\caption{\normalsize Generated flow features}
					\label{table:GeneratedFlowFeatures}\\
					
					\hline &Feature&Description\\ \hline
					
					1&TnBPSrcIP&Total Number of bytes per source IP\\ \hline
					2&TnBPDstIP&Total Number of bytes per Destination IP.\\ \hline
					3&TnP\textunderscore PSrcIP&Total Number of packets per source IP.\\ \hline
					4&TnP\textunderscore PDstIP&Total Number of packets per Destination IP.\\ \hline
					5&TnP\textunderscore PerProto&Total Number of packets per protocol.\\ \hline
					6&TnP\textunderscore Per\textunderscore Dport&Total Number of packets per dport\\ \hline
					7&AR\textunderscore P\textunderscore Proto\textunderscore P\textunderscore SrcIP&Average rate per protocol per Source IP. (calculated by pkts/dur)\\ \hline
					8&AR\textunderscore P\textunderscore Proto\textunderscore P\textunderscore DstIP&Average rate per protocol per Destination IP.\\ \hline
					9&N\textunderscore IN\textunderscore Conn\textunderscore P\textunderscore SrcIP&Number of inbound connections per source IP.\\ \hline
					10&N\textunderscore IN\textunderscore Conn\textunderscore P\textunderscore DstIP&Number of inbound connections per destination IP.\\ \hline
					11&AR\textunderscore P\textunderscore Proto\textunderscore P\textunderscore Sport&Average rate per protocol per sport\\ \hline
					12&AR\textunderscore P\textunderscore Proto\textunderscore P\textunderscore Dport&Average rate per protocol per dport\\ \hline
					13&Pkts\textunderscore P\textunderscore State\textunderscore P\textunderscore Protocol\textunderscore P\textunderscore DestIP&Number of packets grouped by state of flows and protocols per destination IP.\\ \hline
					14&Pkts\textunderscore P\textunderscore State\textunderscore P\textunderscore Protocol\textunderscore P\textunderscore SrcIP&Number of packets grouped by state of flows and protocols per source IP.\\ \hline
					
				\end{longtable}
			\end{small}
		\end{center}
	\section{Benign and Botnet scenarios}
		\subsection{Benign Scenarios}
		In the testbed environment, we design a typical smart-home configuration. Initially, five smart devices were simulated and operated locally. We employed the Node-Red toot to connect smart devices and the corresponding Cloud infrastructure for generating normal/benign network traffic. Moreover, the Ostinato tool was also utilized to generate a huge amount of normal traffic between the VMs, like network production systems.
		\newline\newline
		The configuration of the VMs and utilized platforms represents a realistic smart-home network, as the five IoT devices: 1) Smart Refrigerator, 2) Smart Garage door, 3) Weather Monitoring System, 4) Smart Lights, and 5) Smart thermostat, could be deployed in smart homes. Also, the generated messages were transferred to a Cloud Service provider (AWS) using the MQTT protocol. The statistics of normal traffic included in the dataset are shown in Table \ref{table:StatisticsOfNormalInstancesIncludedInBot-IoTDataset}.
		
			\begin{center}
			\begin{small}
				\begin{longtable}{|p{0.22\textwidth}|p{0.22\textwidth}|}
					\caption{\normalsize Statistics of normal instances included in Bot-IoT dataset}
					\label{table:StatisticsOfNormalInstancesIncludedInBot-IoTDataset}\\
					
					\hline Protocol&Number\\ \hline
					UDP&7225\\ \hline
					TCP&1750\\ \hline
					ARP&468\\ \hline
					IPV6-ICMP&88\\ \hline
					ICMP&9\\ \hline
					IGMP&2\\ \hline
					RARP&1\\ \hline
					Total&9543\\ \hline
					
				\end{longtable}
			\end{small}
		\end{center}
		\subsection{Botnet scenarios}
		As previously mentioned, we used four Kali Linux VMs to launch cyber-attacks in parallel for implementing different botnet scenarios, as depicted in Figure 1. The cyber-attacks and their tools considered in the Bot-IoT dataset are described as follows:
		\begin{itemize}
			\item \textbf{Probing attacks} \cite{paliwal2012denial}\cite{bartlett2007understanding}\cite{hoque2014network}\cite{lyon2009nmap}- are malicious activities that gather information about victims through scanning remote systems, also so-called, fingerprinting \cite{paliwal2012denial}\cite{hoque2014network}. The probing types included in the dataset are further discussed below.

			Probing can be split into subcategories, first based on actions performed during probing and second based on the information gathering target. First, according to the actions performed, probing can be split into passive and active probing \cite{bartlett2007understanding}\cite{hoque2014network}. During passive probing, an attacker simply captures any and all available packets in the network, thus operating in a stealthy manner \cite{bartlett2007understanding}\cite{hoque2014network}. On the other hand, during an active probe, an attacker generates network traffic, targeting the system, and records the responses, comparing them with known responses which allow them to make inferences about services and OS \cite{bartlett2007understanding}\cite{hoque2014network}. With regards to the goal of the probe, there are two major subcategories, OS and service fingerprinting. In OS fingerprinting, a scanner gathers information about the remote system’s OS by comparing its responses to pre-existing ones, or based on differences in TCP/IP stack implementations. In service fingerprinting (scanning), a scanner identifies the services which run behind the system’s ports (0-65535), by sending request packets \cite{hoque2014network}. Here, we will be using active scans, as passive scans produce close to zero amount of generated traffic.
				\begin{itemize}
					\item Port scanning: we used the Nmap and Hping3 tools in order to perform a number of different types of port scans. The Nmap \cite{lyon2009nmap} was launched to scan open services and protocols of the VMs, for example, \textit{nmap -sT 192.168.100.3}, where \textit{sT} issues a complete TCP connection and \textit{192.168.100.3} is the IP of the Ubuntu server. Likewise, Hping3 \cite{2018c} was employed to perform similar port scans with an example being, \textit{hping3 –S –scan 1-1000 192.168.100.3}, where \textit{S} issues a SYN scan and \textit{scan} dictates the port numbers to be scanned.
					\item OS fingerprinting: we used the Nmap and Xprobe2 tools to launch different types of OS fingerprint scans. The Nmap \cite{lyon2009nmap} tool was used to identify the OS of our target VMs with different options, for example, \textit{nmap -sV -T5 -PO -O 192.168.100.3}, where \textit{sV} specifies a Syn scan, \textit{T5} that the scan is as overt as possible, \textit{PO} to include IP protocol ping packets and \textit{O} to enable OS scanning. The Xprobe2 \cite{2018d} tool was used in conjunction with Nmap. While performing our scans, we used the default operations of Xprobe2, with no options specified.
				\end{itemize}
			\item \textbf{Denial of Service} \cite{behal2017detection}\cite{doshi2018machine}\cite{kolias2017ddos}\cite{zargar2013survey}- are malicious activities that attempt to disrupt a service, thus making it unavailable to legitimate users. The DDoS and DoS attack types included in the dataset are described as follows:
			Distributed Denial of Service (DDoS) and Denial of Service (DoS) attacks are performed by a group of compromised machines called Bots, and target a remote machine, usually a server\cite{behal2017detection}\cite{doshi2018machine}\cite{kolias2017ddos}. The purpose of such attacks is the disruption of services accessible by legitimate users. These attacks can be classified, based on their attack methodology. Two such groups are volumetric and protocol-based DDoS/DoS attacks\cite{zargar2013survey}. Volumetric attacks generate a great number of network traffic, which either force the victim to process through these attack-generated requests, or cause the machine to crash, thus making the provided service unavailable. Protocol-based attacks abuse the mechanics of Internet protocols, which cause CPU and memory resources to be depleted, thus render the targeted machine unable to respond to requests. In our attack scenarios, we performed both DDoS and DoS, and used the following protocols: TCP, UDP and HTTP.			
				\begin{itemize}
					\item DDoS, DoS: We used the tool Hping3 \cite{2018c} for both DDoS, DoS for TCP and UDP, for example, \textit{hping3 --syn --flood -d 100 -p 80 192.168.100.3} where \textit{syn} indicates a SYN TCP attack, \textit{flood} indicates packets are sent as fast as possible, \textit{d} specifies packet body size, \textit{p} sets the targeted port.
					For HTTP DDoS and DoS attacks, we used the Golden-eye tool, one example being \textit{goldeneye.py http://192.168.100.3:80 -m post -s 75 -w 1}, with \textit{http://192.168.100.3:80} indicating the IP address of Ubuntu Server and the targeting Port number, \textit{m} setting the method as post, \textit{s} setting number of sockets, \textit{w} specifying number of concurrent workers. 
					
				\end{itemize}
			\item \textbf{Information Theft} \cite{tankard2011advanced}\cite{jesudoss2014survey}- is a group of attacks where an adversary seeks to compromise the security of a machine in order to obtain sensitive data. The information theft attack types included in the dataset are described as follows:
			
			Information theft attacks can be split into subcategories, based on the target of the attack. The first subcategory is data theft. During data theft attacks, an adversary targets a remote machine and attempts to compromise it, thus gaining unauthorized access to data, which can be downloaded to the remote attacking machine. The second subcategory is keylogging. In keylogging activities, an adversary compromises a remote host in order to record a user’s keystrokes, potentially stealing sensitive credentials. Attackers usually employ Advanced Persistent Threat (APT) methodology in conjunction with information Theft attacks, in order to maximize their attack’s efficiency \cite{tankard2011advanced}.
				\begin{itemize}
					\item Data theft: we used the Metasploit framework \cite{2018h} to exploit weaknesses in the target machines. For Windows 7 we exploited the SMB eternal blue vulnerability, while for Ubuntu Server we took advantage of weak administrator credentials. Post exploitation, we set-up a reverse meterpeter tcp connection through which exfiltration of entire directories became possible.
					\item Keylogging: we used the Metasploit framework \cite{2018h} to exploit the same weaknesses we used during Data theft. For Windows 7, meterpeter provided adequate software to perform keylogging, although the unpredictable shutdowns of the exploit itself rendered a collection of keystrokes impossible. For Ubuntu Server, we used the logkeys \cite{2018f} software to record keystrokes. Initially, a dictionary attack was launched through Hudra \cite{2018g} on the SSH service. Then, through Metasploit, an ssh connection was established which was later upgraded to a sudo meterpeter connection, allowing us to access the logkeys software, record keystrokes in the compromised host and then download the recordings.
				\end{itemize}
		\end{itemize}
	The statistics of attacks involved in the dataset are described in Table \ref{table:StatisticsOfAttacksInIoT-BotDataset}

	\begin{center}

		\begin{small}
			\begin{longtable}{|l|l|l|l|l|}
				\caption{\normalsize Statistics of attacks in IoT-Bot dataset}
				\label{table:StatisticsOfAttacksInIoT-BotDataset}\\\hline

				\multirow{2}{*}{Information gathering}&\multicolumn{2}{|l|}{Service scanning}&nmap, hping3&1463364 \\ \cline{2-5}
				& \multicolumn{2}{|l|}{OS Fingerprinting}&nmap, xprobe2&358275\\ \cline{1-5}
				\multirow{6}{*}{Denial of Service} 
				& \multirow{3}{*}{DDoS} & TCP & hping3&19547603 \\ \cline{3-5}
				& & UDP & hping3&18965106 \\ \cline{3-5}
				& & HTTP & golden-eye&19771 \\ \cline{2-5}
				& \multirow{3}{*}{DoS} & TCP & hping3&12315997 \\ \cline{3-5}
				& & UDP & hping3&20659491 \\ \cline{3-5}
				& & HTTP & golden-eye&29706 \\ \cline{1-5}
				\multirow{2}{*}{Information theft}&\multicolumn{2}{|l|}{Keylogging}&Metasploit&1469 \\ \cline{2-5}
				& \multicolumn{2}{|l|}{Data theft}&Metasploit&118\\ \cline{1-5}
				\multicolumn{1}{|l|}{Total}&\multicolumn{3}{|l|}{}&73360900 \\ \cline{1-5} 
			\end{longtable}
		\end{small}
	\end{center}
	\section{Statistics and machine learning methods}
	This section describes a theoretical background of statistical measures for evaluating optimal features and machine learning for forensically identifying cyber-attacks.
		\subsection{Statistical analysis techniques}
		\begin{itemize}
			\item \textit{Pearson Correlation Coefficient:} is used for measuring the linear relationship between the feature set of the Bot-IoT dataset. Its output ranges between [-1,1], and its magnitude indicates the strength of correlation between two feature vectors, and its sign indicates the type of correlation either positive or negative \cite{hall2015pearson}.
			\item \textit{Entropy:} depicts the uncertainty or disorder between features \cite{zheng2011feature} of the Bot-IoT dataset. By definition, high values of entropy equate to high uncertainty.
			\newline

				\begin{equation}
				\label{ShannonJointEntropy}	
					-\sum_{x}\sum_{y}(P(x,y)*\log P(x,y))
				\end{equation}

			Based on Equation \ref{ShannonJointEntropy}, the produced entropy values would be greater than or equal to zero H(X,Y)≥0, as such the minimum value of entropy is 0. We calculated the pairwise Shannon Joint Entropy \cite{lesne2014shannon} of all rows, excluding class features, resulting in an n,n table., where n is the number of features.
		\end{itemize}
		\subsection{Machine and Deep Learning analysis techniques}
		Machine and Deep Learning models were used to evaluate the quality of Bot-IoT, when used to train a classifier. The models that were trained were: Support Vector Machine (SVM), Recurrent Neural Network (RNN) and Long-Short Term Memory Recurrent Neural Network (LSTM-RNN).
		\begin{itemize}
			\item \underline{SVM:} is based on the idea that data instances can be viewed as coordinates in an N-dimensional space, with N being the number of features. During training, a hyperplane is sought, that best separates the data into distinct groups (classes) and that maximize the margin. In our work, we made use of an SVM Classifier with a linear kernel \cite{meyer2001support}.
			\item \underline{RNN:} incorporates a form of memory in its structure \cite{grossberg2013recurrent}. The output of an RNN during an iteration, depends both on the input at any given time, and the output of the Hidden state of the previous iteration. What makes RNN stand out, is that contrary to other NNs, its output depends on both the current input as-well-as previous outputs, making it ideal for processing temporal data, such as the ones present in our dataset. Usual applications of RNNs include machine translation, speech recognition, generating Image descriptions.
			\item \underline{LSTM:} is a special kind of Recurring Neural Network, where a collection of internal cells are tasked with maintaining a sort of memory which makes LSTMs ideal for finding temporally distant associations \cite{greff2017lstm}. LSTM improve on RNN's \lq vanishing gradient\rq and \lq exploding gradient\rq problems, by incorporating a \lq memory cell\rq which handles updates in the model's memory. This improvement renders LSTMs ideal for learning long-term dependencies in data.
		\end{itemize}
	\section{Experimental Results and Discussion}
		\subsection{Pre-processing steps of Bot-IoT dataset}
		In order to extract the dataset from its MySQL tabular form, which contains the combined records from all the subsequent tables, where the labelling process took place, we introduced an auto-incrementing feature named \lq pkSeqID\rq, and then employed \textit{“select * from IoT\textunderscore Dataset\textunderscore UNSW\textunderscore 2018 into outfile '/path/to/file.csv'” Fields terminated by ',' Lines terminated by ' \textbackslash n';}, to extract the dataset into csv form. Doing so enabled us to easily process the data with various Python modules and also makes distributing the dataset easier, as csv format is widely used for such purposes.
		\newline\newline		
		Furthermore, considering that the generated dataset is very large (more than 72.000.000 records and at 16.7 GB for CSV, with 69.3 GB pcap), , it made handling the data very cumbersome. As such, we extracted 5\% of the original dataset via the use of select MySQL queries similar to the ones mentioned previously. The extracted 5\%, which we will refer to as the training and testing sets for the rest of the paper, is comprised of 4 files of approximately 0.78 GB  total size, and about 3 million records.
		\newline\newline
		Additionally, it became necessary at some point of the evaluation process, to introduce discrete copies of numerical features. To do so, we grouped the numeric values into 5 bins of equal size, and later used the mathematical representation of the produced sets \lq(min, max)\rq of each bin as the new discrete value for each corresponding numeric in the dataset.\newline\newline
		Moreover, due to the existence of certain protocols (ARP), source and destination port number values were missing (not applicable), as such, these values were set to -1, which is an invalid port number, again for the purpose of evaluation of the dataset.
		We converted the categorical feature values in the dataset into consecutive numeric values for easily applying statistical methods. For example, the state attribute has some categorical values such as \lq RST\rq, \lq CON\rq, and \lq REQ\rq that were mapped into \lq 1\rq , \lq 2\rq and \lq 3\rq.
		\newline\newline
		Moreover, normalization was applied in order to scale the data into a specific range, such as [0,1], without changing the normality of data behaviors. This step helps statistical models and deep learning methods to converge and achieve their objectives by addressing local optimum challenges. We performed a Min-Max transformation on our data, according to the following formula:
		
		\begin{equation}
		\label{min-max transformation}	
		x_{i}^{\acute{}}=(x_{i}-x_{min})* \frac{(b-a)}{(x_{max}-x_{min})}+a
		\end{equation}
		Where xmax and xmin are the initial max an min values from the original set, b and a are the new max and min set values and 〖$x_{i}^{'}\in[a,b]$. For our purposes, a=0 and b=1, making the new set [0,1].In order to measure the performance of the trained models, corresponding confusion matrices were generated, along with a collection of metrics, as given in Table \ref{table:MachineLearningEvaluationMetrics}.
		\begin{center}
			\begin{small}
				\begin{longtable}{|p{0.22\textwidth}|p{0.32\textwidth}|}
					\caption{\normalsize Machine Learning evaluation metrics}
					\label{table:MachineLearningEvaluationMetrics}\\ \hline
					\Large Accuracy& \Large$ACC=\frac{TP}{TP+FP}$\\ \hline
					\Large Precision&\Large$PPV=\frac{TP}{TP+FP}$\\ \hline
					\Large Recall&\Large$TPR=\frac{TP}{TP+FN}$\\ \hline
					\Large Fall-out&\Large$FPR=\frac{FP}{FP+TN}$\\ \hline
										
				\end{longtable}
			\end{small}
		\end{center}
		
		\subsection{Unsupervised Attribute evaluations}
		Initially, we followed a filter method feature selection. Such methods rely on statistical techniques to evaluate the quality of features rather than the optimization of ML models. The idea behind these evaluations, is to identify any features that are highly correlated and reduce the dimensionality for improving the performances of machine learning models.
			\subsubsection{Correlation Coefficient}
			In order to calculate the correlation coefficient between the dataset’s features, we developed a code in Python to rank the attribute strengths into a range of [-1, 1]. After calculating the Correlation Coefficient Matrix, we computed the average Correlation for each feature under scrutiny, thus gaining the “average correlation”. The idea is that the features with the lowest Correlation Coefficient Average would introduce less ambiguity in our dataset. In Table \ref{table:AverageCorrelationCoefficientScores}, the features are represented in order from the lowest produced average correlation to the highest.
			\begin{center}
				\begin{small}
					\begin{longtable}{|l|l|l|l|l|l|l|}
						\caption{\normalsize Average Correlation Coefficient scores}
						\label{table:AverageCorrelationCoefficientScores}\\ \hline
						\textbf{Features}   & srate& drate& seq& rate& dur& min\\\hline
						\textbf{Average CC} & -0.00061 & -0.00502 & 0.019941 & -0.03508 & 0.051058 & 0.070493\\\hline
						\textbf{Featyres (1)} & stddev   & stime    & ltime    & flgs\_number & state\_number & mean\\\hline
						\textbf{Average CC (1)} & 0.077707 & 0.102708 & 0.102716 & -0.10681     & 0.106846      & 0.170598\\\hline
						\textbf{Features (2)}  & max      & dpkts    & dbytes  & sbytes   & spkts    & bytes\\\hline
						\textbf{Average CC (2)}& 0.176318 & 0.237771 & 0.23879 & 0.261992 & 0.265161 & 0.277309\\\hline
						\textbf{Features (3)} &pkts&Sum\\\cline{1-3}
						\textbf{Average CC (3)} &0.284521&0.288727\\\cline{1-3}
					\end{longtable}
				\end{small}
			\end{center}
		
			\subsubsection{Entropy}
			In order to calculate the joint entropy between our features, we generated Python code which traversed the loaded CSV files and calculated the subsequent sums of joint probability times the base 2 logarithm for that probability, as given in Equation 1. It was due to this measure that we performed the discretization that we mentioned at the beginning of this section. 
			\newline\newline
			Contrary to the Correlation Coefficient, Entropy depicts disorder in data, and as such, higher values indicate higher disorder (or randomness), which means that our features do not share much information and thus introduce less ambiguity in our dataset. Thus, we again produce a score value per feature, through calculating the average Joint Entropy, as depicted in the following table from higher to lower average Entropy.
			\begin{center}
				\begin{small}
					\begin{longtable}{|l|l|l|l|l|l|l|}
						\caption{\normalsize Average Joint Entropy scores for BoT-IoT features}
						\label{table:AverageJointEntropyScoresForBoT-IoTFeatures}\\ \hline
						\textbf{Features}& seq          & mean     & stddev   & max      & Min      & state\_number \\\hline
						\textbf{Average JE}& 2.833223     & 2.465698 & 2.423786 & 2.088409 & 2.015226 & 1.963397      \\\hline
						\textbf{Features (1)}& flgs\_number & stime    & ltime    & dur      & rate     & sum           \\\hline
						\textbf{Average JE (1)}& 1.68267      & 0.798997 & 0.798997 & 0.665417 & 0.664154 & 0.661829      \\\hline
						\textbf{Features (2)}& spkts        & pkts     & sbytes   & bytes    & srate    & drate         \\\hline
						\textbf{Average JE (2)}& 0.661812     & 0.661783 & 0.661772 & 0.661768 & 0.661734 & 0.66172       \\\hline
						\textbf{Features (3)}& dbytes       & dpkts \\\cline{1-3}
						\textbf{Average JE (3)}& 0.661718     & 0.661718\\\cline{1-3}              
					\end{longtable}
				\end{small}
			\end{center}
			
			\subsubsection{Extraction of the 10 Best features}
\begin{figure}[H]
				\centering
				\includegraphics[width=\textwidth]{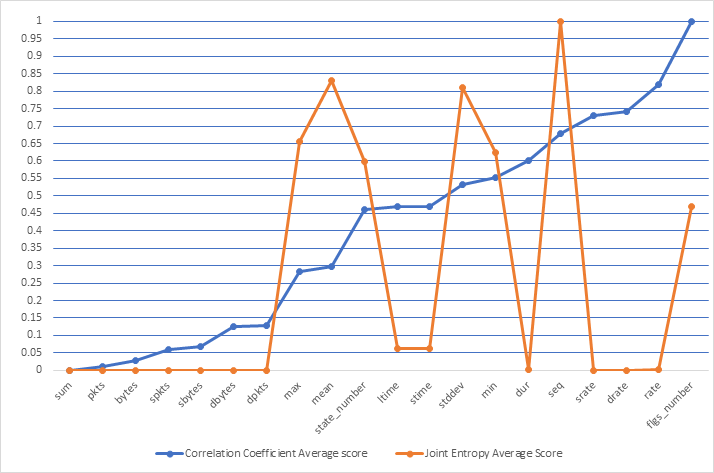}
				\caption{Graph representation of features}
				\label{fig:Graph_representation_of_features}
			\end{figure}

			A direct comparison of both Entropy and Correlation scores is given in Figure ~\ref{fig:Graph_representation_of_features}. A feature will be considered ideal for our dataset, if its Entropy score is large enough and its Correlation Score low enough. That would mean that that feature does not carry any redundant information that is shared with other features and that they are as unrelated with each other as possible. In order to compare the averages of these different statistical measures, score values were normalized in the range [0,1].
			\newline\newline
			Considering that higher Correlation Coefficient values indicate highly correlated features, which is something we wanted to remove from our dataset, after performing the Min-Max transformation, we inverted the results $(1-y_{i})$, so as to bring the CC average scores in the same format as the Joint Entropy score (where higher values translates to higher randomness between features).
			we compared the new mapped values of Correlation Coefficient and Joint Entropy in order to extract a subset of 10 features which, overall, had the best scores in both statistical measures. As such, we identified toe following best 10 features: srate, drate, rate, max, state\textunderscore number, mean, min, stddev, flgs\textunderscore number, seq. Having completed the unsupervised evaluation process, we will further evaluate the worthiness of the final 10 features in the section \lq supervised evaluation\rq.
			\subsubsection{Secondary evaluation of features}
			In this section, we further evaluated the relationship between the independent features. For this stage, we employed similar tactics to the previous section of this paper, that is, we calculated average scores based on the Pearson Correlation Matrix (Triangle) and Shannon Joint Entropy, for the 10 best features that were identified previously, combined with the 14 generated features, and then extracted the 10 best features from that group. Following are the Joined Entropy and Correlation Coefficient matrices and plots of their average scores.
			\begin{center}
				\begin{small}			
					\begin{longtable}{|p{0.4\textwidth}|l|l|}
						\caption{\normalsize Joint Entropy and Correlation Coefficnent average scores}
						\label{table:Joint Entropy and Correlation Coefficnent average scores}
						\\\hline
						\textbf{Features}      & \textbf{Average CC} & \textbf{Average JE} \\ \hline
						AR\_P\_Proto\_P\_Dport & 0.071521            & 0.637848            \\ \hline
						AR\_P\_Proto\_P\_DstIP & 0.034548            & 0.636399            \\ \hline
						AR\_P\_Proto\_P\_Sport & 0.070876            & 0.636738            \\ \hline
						AR\_P\_Proto\_P\_SrcIP & 0.034062            & 0.636084            \\ \hline
						drate                  & 0.004333            & 0.635576            \\ \hline
						flgs\_number           & -0.13459            & 1.680341            \\ \hline
						max                    & 0.18572             & 2.104321            \\ \hline
						mean                   & 0.192966            & 2.490935            \\ \hline
						min                    & 0.093812            & 2.022061            \\ \hline
						N\_IN\_Conn\_P\_DstIP  & 0.07164             & 1.430813            \\ \hline
						N\_IN\_Conn\_P\_SrcIP  & 0.077322            & 2.067795            \\ \hline
						Pkts\_P\_State\_P\_Protocol\_P\_SrcIP    & 0.263217            & 0.635667            \\ \hline
						Pkts\_P\_State\_P\_Protocol\_P\_DestIP    & 0.266336            & 0.635636            \\ \hline
						    
						rate                   & 0.074073            & 0.638024            \\ \hline
						seq                    & -0.02858            & 2.835564            \\ \hline
						srate                  & 0.005537            & 0.63559             \\ \hline
						state\_number          & 0.122872            & 1.973091            \\ \hline
						stddev                 & 0.072967            & 2.440526            \\ \hline
						TnBPDstIP              & 0.168677            & 0.635674            \\ \hline
						TnBPSrcIP              & 0.161545            & 0.635689            \\ \hline
						TnP\_PDstIP            & 0.271537            & 0.635701            \\ \hline
						TnP\_Per\_Dport        & 0.242596            & 0.63563             \\ \hline
						TnP\_PerProto          & 0.130232            & 0.636924            \\ \hline
						TnP\_PSrcIP            & 0.268818            & 0.635686            \\ \hline
					\end{longtable}
				\end{small}
		\end{center}
		We then mapped the average scores in the set [0,1] and plotted the values, in order to identify the 10 best features.
		\begin{figure}[H]
			\centering
			\includegraphics[width=\textwidth]{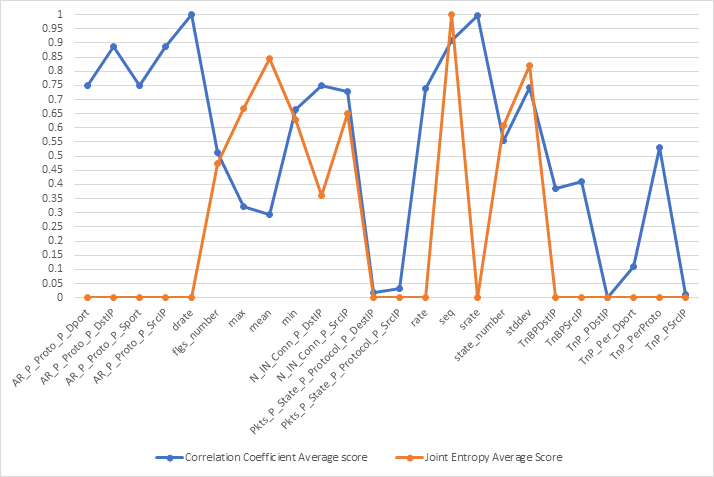}
			\caption{Graph representation of features' scores}
			\label{fig:Graph_representation_of_features_scores}
		\end{figure} 
		By observing Figure ~\ref{fig:Graph_representation_of_features_scores}, we can determine that the 10 best features, that is, the 10 features with the highest combined average Correlation Coefficient and Joint Entropy are: \textit{seq, stddev, N\textunderscore IN\textunderscore Conn\textunderscore P\textunderscore SrcIP, min, state\textunderscore number, mean, N\textunderscore IN\textunderscore Conn\textunderscore P\textunderscore DstIP, drate, srate, max}.
		\subsection{Supervised evaluation}
		After the 10-best features were extracted, we employed supervised learning techniques, to evaluate the quality of the dataset. Such methods rely on a model which is trained on the labeled data, and then is capable of classifying new unlabeled instances.
			\subsubsection{Three classifiers for evaluations}
				\setcounter{secnumdepth}{4}
				Following the selection of the 10 best features, we applied three predictors on our data, and assessed their accuracy, in order to further eliminate any superfluous features. Specifically, the predictors we chose were a \textbf{S}upport \textbf{V}ector \textbf{M}achine (SVM), a \textbf{R}ecurring \textbf{N}eural \textbf{N}etwork (RNN)and a \textbf{L}ong \textbf{S}hort-\textbf{T}erm \textbf{M}emory RNN (LSTM-RNN).
				
				\paragraph{SVM}
				\setcounter{secnumdepth}{4}\mbox{}
				
				The SVM model that was trained, was a linear Support Vector Classifier. The parameters of this classifier were penalty parameter (C=1), 4-fold cross-validation and a number of max iterations equal to 100000 on the final 10-best features. Similar settings were selected for the dataset version comprised of all available features, with the only difference being that max iterations were set to 1000.
				\newline\newline The aforementioned setting was practically adjusted to measure the best performance of the SVM model. Initially, the SVM classifier was trained with default parameters, but it was later observed that by increasing the max iteration number, particularly for the second (all features included) model caused a longer training time. With regards to the number of folds, we observed a loss of accuracy when a higher number of folds was chosen.
								
				\begin{center}
					\begin{small}
						\begin{table}[h]
							\caption{\normalsize Confusion matrices of SVM models.
								(10-best feature model on the left, full-feature model on the right).}
								
							\begin{minipage}{.5\linewidth}
							\addtocounter{table}{-1}					
								\begin{longtable}{p{.3\textwidth}|p{.28\textwidth}|l}
										True\textbackslash{}Predict & Normal (0) & \multicolumn{1}{l}{Attack (1)} \\ \hline
										Normal (0) & 477 & 0  \\ \hline
										Attack (1) & 426550 & 3241495                        
								
								\end{longtable}
							\end{minipage}
							\begin{minipage}{.5\linewidth}
								\addtocounter{table}{-1}					
								\begin{longtable}{||p{.3\textwidth}|p{.28\textwidth}|l}
									True\textbackslash{}Predict & Normal (0) & \multicolumn{1}{l}{Attack (1)} \\ \hline
									Normal (0) & 64 &413 \\ \hline
									Attack (1) & 0 & 3668045                        
									
								\end{longtable}
							\end{minipage}
							
						\end{table}
					\end{small}
				\end{center}
				
				\begin{figure}[H]
					\centering
					\includegraphics[width=\textwidth]{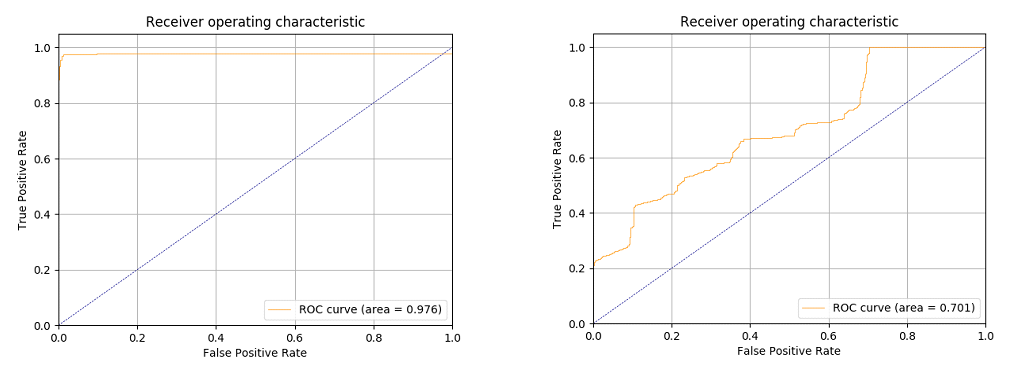}
					\caption{ROC curve for SVM models.(10-best feature model on the left,full-feature model on the right).}
					\label{fig:ROC_Curve_for_SVMs}
				\end{figure}

				\paragraph{LSTM}
				\setcounter{secnumdepth}{4}\mbox{}
								
\begin{figure}[H]
					\centering
					\includegraphics[width=\textwidth]{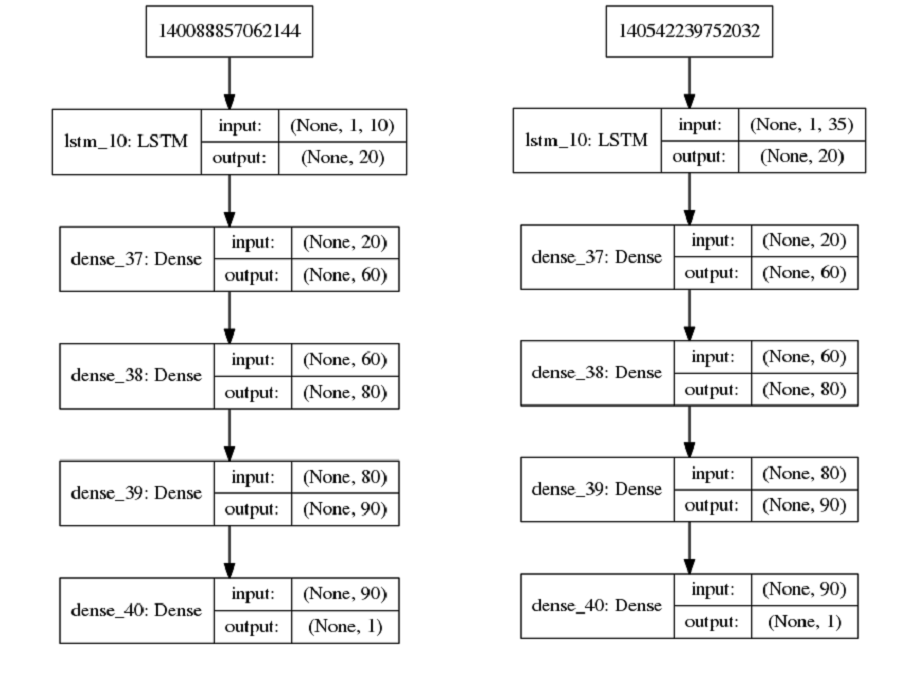}
					\caption{Structure of LSTM model (in layers).
						(10-best feature model on the left, full-feature model on the right)}
					\label{fig:Structure_of_LSTM_model}
				\end{figure}

				The LSTM models were defined to have 1 input layer with the number of neurons equal to the number of input features, two hidden layers and an output layer. For the 10-best features dataset, the Model was trained in 4 epochs with batch size 100. The neural network was comprised of 10 input neurons (in the first layer, the same number as the features), intermediate (hidden) layers with 20, 60,80, 90 neurons and 1 output neuron for the binary classification.\newline\newline
				For our full-feature dataset, the model was trained in 4 epochs with a batch size of 100 records. The network had a 35-neuron input layer (again, the same number as features of the dataset), with similar hidden layers to the model we used to train the 10-best feature version of our dataset (20,60,80,90 neurons) and 1 output neuron for the binary classification.\newline\newline
				We initially tested the model with a batch size of 1000, but due to poor performance, we sought a different value. It was observed, that choosing a batch size of 100 records, the specificity of the model was improved. 
				In both cases, for the input and hidden layers, the activation function that was used was \lq tanh\rq, while the output layer activation function was \lq sigmoid\rq. Both tanh and sigmoid activation functions are often used for building Neural Networks, with sigmoid being an ideal choice for binary classification, as its output is within the [0,1] set. Bellow the structure of the LSTM model can be viewed in Figure ~\ref{fig:Structure_of_LSTM_model}. 
				
				Next, we present the training times for both LSTM models, the confusion matrices followed by four metrics.
				\begin{center}
					\begin{small}
						\begin{table}[H]
							\caption{\normalsize Confusion matrices of LSTM models.
								(10-best feature model on the left, full-feature model on the right).}
							
							\begin{minipage}{.5\linewidth}
								\addtocounter{table}{-1}
								\begin{longtable}{p{.3\textwidth}|p{.28\textwidth}|l}
									True\textbackslash{}Predict & Normal (0) & \multicolumn{1}{l}{Attack (1)} \\ \hline
									Normal (0) & 149 & 328  \\ \hline
									Attack (1) & 9139 & 3658906                        
									
								\end{longtable}
							\end{minipage}
							\begin{minipage}{.5\linewidth}
								\addtocounter{table}{-1}
								\begin{longtable}{||p{.3\textwidth}|p{.28\textwidth}|l}
									True\textbackslash{}Predict & Normal (0) & \multicolumn{1}{l}{Attack (1)} \\ \hline
									Normal (0) & 430 &47 \\ \hline
									Attack (1) & 71221 & 3596824                        
									
								\end{longtable}
							\end{minipage}
							
						\end{table}
					\end{small}
				\end{center}
				\begin{figure}[H]
					\centering
					\includegraphics[width=\textwidth]{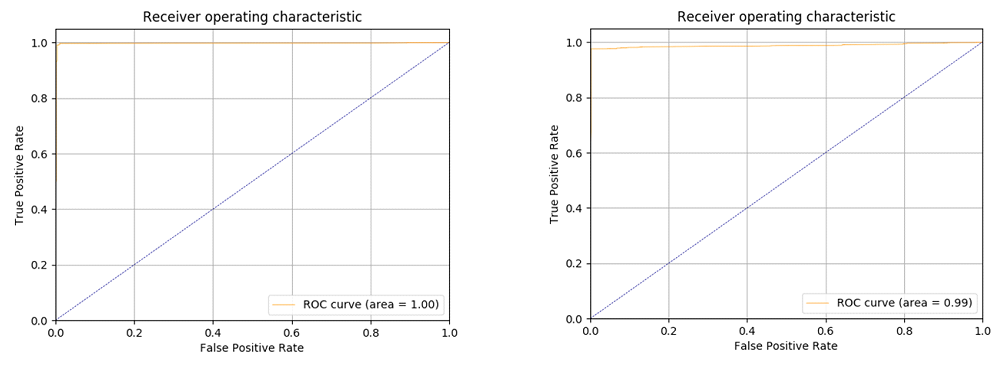}
					\caption{ROC curve for LSTM models. (10-best feature model on the left, full-feature model on the right)}
					\label{fig:ROC_Curve_for_LSTM}
				\end{figure}
				\paragraph{RNN}\mbox{}\newline
				The RNN models were defined to have 1 input layer with the number of neurons equal to the number of input features, two hidden layers and an output layer. For the 10-best features dataset, the Model was trained in 4 epochs (batch size 100), had 10 input neurons (in the first layer, the same number as the features), with hidden layers similar to the ones in the LSTM models, and 1 output neuron for the binary classification.As with LSTM-RNN, the parameters were chosen through experimentation. Higher values of batch size affected the model’s specificity, as such we experimented with lower values.\newline\newline
				For our full-feature dataset, the model was trained in 4 epochs (batch size 100), and had a 35-neuron input layer, same number and consistency of hidden layers as with the 10-best feature model and 1 output neuron for the binary classification.
				In both cases, for the input and hidden layers, the activation function that was used was \lq tanh\rq, while the output layer activation function was \lq sigmoid\rq. As mentioned previously, the sigmoid function is ideal for output layers for a binary classification problem, as its output is within the [0,1] set. Bellow the structure of the LSTM model can be viewed in Figure ~\ref{fig:Structure_of_RNN_model}.
				\begin{figure}[H]
					\centering
					\includegraphics[width=\textwidth]{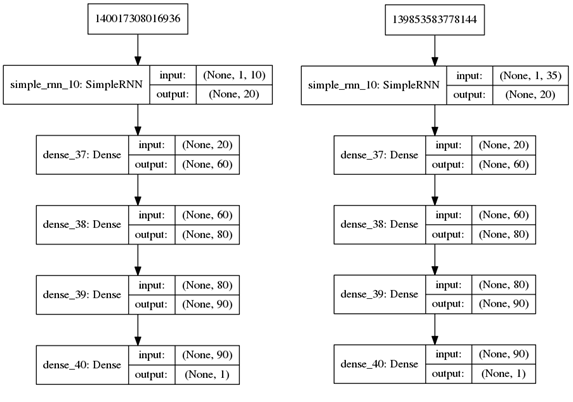}
					\caption{Structure of RNN model (in layers). (10-best feature model on the left, full-feature model on the right)}
					\label{fig:Structure_of_RNN_model}
				\end{figure}

				Next, we present the training times for both LSTM models, the confusion matrices followed by four metrics.
				
				\begin{center}
					\begin{small}
						\begin{table}[H]
							\caption{\normalsize Confusion matrices of RNN models.
								(10-best feature model on the left, full-feature model on the right).}
							
							\begin{minipage}{.5\linewidth}
								\begin{longtable}{p{.3\textwidth}|p{.28\textwidth}|l}
									True\textbackslash{}Predict & Normal (0) & \multicolumn{1}{l}{Attack (1)} \\ \hline
									Normal (0) & 127 & 350  \\ \hline
									Attack (1) & 9171 & 3658874                        
									
								\end{longtable}
							\end{minipage}
							\begin{minipage}{.5\linewidth}	
								\addtocounter{table}{-1}				
								\begin{longtable}{||p{.3\textwidth}|p{.28\textwidth}|l}
									True\textbackslash{}Predict & Normal (0) & \multicolumn{1}{l}{Attack (1)} \\ \hline
									Normal (0) & 379 &98 \\ \hline
									Attack (1) & 76718 & 3591327                        
									
								\end{longtable}
							\end{minipage}
							
						\end{table}
					\end{small}
				\end{center}
			    \begin{figure}[H]
			    	\centering
			    	\includegraphics[width=\textwidth]{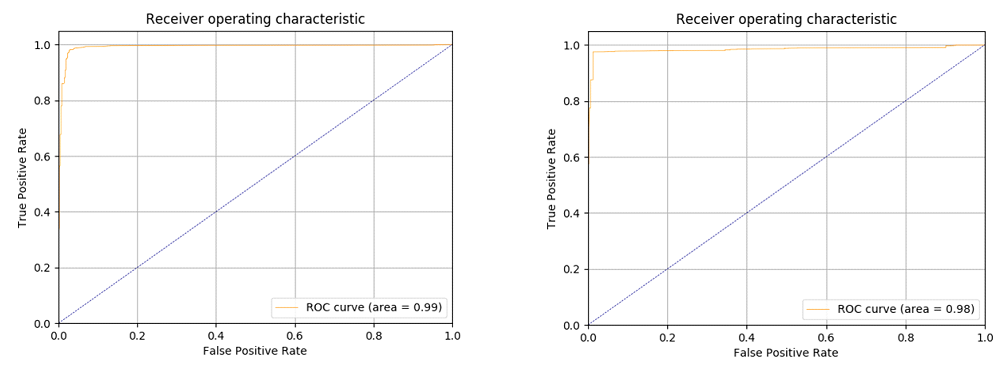}
			    	\caption{ROC curve for RNN models.
			    		(10-best feature model on the left, full-feature model on the right)}
			    	\label{fig:ROC_Curve_for_RNN_models}
			    \end{figure}
				Finally, we evaluated a simple Deep RNN network’s capabilities of distinguishing between normal traffic and each of the attacks that are described in the dataset separately.\newline\newline
				The network had 10 input neurons, and 8 hidden layers comprised of 30, 40, 40, 60, 80, 80, 100 neurons each. Finally, the output layer had 1 neuron for binary classification. The resulting confusion matrices can be viewed in Table ~\ref{table:RNN_Confusion_Matrices_Sub_Cat_Eval}
				
				In the following table (~\ref{table:Accuracy,Precision,RecallFall-outAndTrainingTimeMetrics}), we present the four metrics Accuracy, Precision, Recall, Fall-out along with Training Time of all the models that were trained.
			
			\begin{turn}{90}
				\begin{minipage}{\textheight}
					\addtocounter{table}{-1}
					\begin{table}[H]
						\caption{ Accuracy, Precision, Recall Fall-out and training time Metrics. (In tables(b) and (c), the model was an RNN. Training time is in seconds)}
						\label{table:Accuracy,Precision,RecallFall-outAndTrainingTimeMetrics}

				$$\vbox{
					\offinterlineskip
					\halign{
						\strut
						\vrule height1ex depth1ex width0px #
						&\vrule\kern3pt #\hfil\kern3pt\vrule
						&\kern3pt #\hfil\kern3pt\vrule
						&\kern3pt #\hfil\kern3pt\vrule
						&\kern3pt #\hfil\kern3pt\vrule
						&\kern3pt #\hfil\kern3pt\vrule
						&\kern3pt #\hfil\kern3pt\vrule
						&\kern3pt #\hfil\kern3pt\vrule
						&\kern3pt #\hfil\kern3pt\vrule
						&\kern3pt #\hfil\kern3pt\vrule
						&\kern3pt #\hfil\kern3pt\vrule
						\cr
						\omit&&\omit\hrulefill&\omit\hrulefill&\omit\hrulefill&\omit\hrulefill&\omit\hrulefill&\omit\hrulefill&\omit\hrulefill&\omit\hrulefill&\omit\hrulefill\cr
						& \omit\kern3pt (a)\hfill\kern3pt \vrule & \multispan2\kern3pt \hfill SVM\hfill\kern3pt \vrule & \multispan4\kern3pt \hfill RNN\hfill\kern3pt \vrule                                                          & \multispan3\kern3pt \hfill LSTM\hfill\kern3pt \vrule                                                                  \cr
						\omit&&\omit\hrulefill&\omit\hrulefill&\omit\hrulefill&\omit\hrulefill&\omit\hrulefill&\omit\hrulefill&\omit\hrulefill&\omit\hrulefill&\omit\hrulefill\cr
						& \omit\kern3pt \hfill\kern3pt \vrule    & 10-best           & Full features                   & \multispan2\kern3pt 10-best\hfill\kern3pt \vrule    & \multispan2\kern3pt Full features\hfill\kern3pt \vrule & \multispan2\kern3pt 10-best\hfill\kern3pt \vrule    & Full features                                                   \cr
						\noalign{\hrule}
						& Accuracy                               & 0.88372702        & 0.99988742                      & \multispan2\kern3pt 0.99740468\hfill\kern3pt \vrule & \multispan2\kern3pt 0.97906078\hfill\kern3pt \vrule    & \multispan2\kern3pt 0.9974194\hfill\kern3pt \vrule  & 0.9805731                                                       \cr
						\noalign{\hrule}
						& Precision                              & 1                 & 0.99988742                      & \multispan2\kern3pt 0.99990435\hfill\kern3pt \vrule & \multispan2\kern3pt 0.99997271\hfill\kern3pt \vrule    & \multispan2\kern3pt 0.99991036\hfill\kern3pt \vrule & 0.99998693                                                      \cr
						\noalign{\hrule}
						& Recall                                 & 0.8837119         & 1                               & \multispan2\kern3pt 0.99749976\hfill\kern3pt \vrule & \multispan2\kern3pt 0.97908477\hfill\kern3pt \vrule    & \multispan2\kern3pt 0.99750848\hfill\kern3pt \vrule & 0.98058339                                                      \cr
						\noalign{\hrule}
						& Fall-out                               & 0                 & 0.86582809                      & \multispan2\kern3pt 0.73375262\hfill\kern3pt \vrule & \multispan2\kern3pt 0.20545073\hfill\kern3pt \vrule    & \multispan2\kern3pt 0.68763103\hfill\kern3pt \vrule & 0.09853249                                                      \cr
						\noalign{\hrule}
						& Training Time                          & 1270.48           & 6636.98                         & \multispan2\kern3pt 8035\hfill\kern3pt \vrule       & \multispan2\kern3pt 6888.08\hfill\kern3pt \vrule       & \multispan2\kern3pt 10482.19\hfill\kern3pt \vrule   & 14073.63                                                        \cr
						\noalign{\hrule}
						& \omit\kern3pt (b)\hfill\kern3pt \vrule & Normal- DDoS HTTP & \multispan2\kern3pt Normal- DDoS TCP\hfill\kern3pt \vrule          & \multispan2\kern3pt Normal  DDoS UDP\hfill\kern3pt \vrule     & \multispan2\kern3pt Normal- DoS HTTP\hfill\kern3pt \vrule          & \multispan2\kern3pt Normal- DoS TCP\hfill\kern3pt \vrule    \cr
						\noalign{\hrule}
						& Accuracy                               & 0.99317872        & \multispan2\kern3pt 0.99991921\hfill\kern3pt \vrule                & \multispan2\kern3pt 0.99999473\hfill\kern3pt \vrule            & \multispan2\kern3pt 0.98063201\hfill\kern3pt \vrule                & \multispan2\kern3pt 0.99993672\hfill\kern3pt \vrule         \cr
						\noalign{\hrule}
						& Precision                              & 0.99295065        & \multispan2\kern3pt 0.99999488\hfill\kern3pt \vrule                & \multispan2\kern3pt 0.99999684\hfill\kern3pt \vrule            & \multispan2\kern3pt 0.98984428\hfill\kern3pt \vrule                & \multispan2\kern3pt 0.99994479\hfill\kern3pt \vrule         \cr
						\noalign{\hrule}
						& Recall                                 & 0.99696663        & \multispan2\kern3pt 0.99992429\hfill\kern3pt \vrule                & \multispan2\kern3pt 0.99999789\hfill\kern3pt \vrule            & \multispan2\kern3pt 0.98451178\hfill\kern3pt \vrule                & \multispan2\kern3pt 0.99999188\hfill\kern3pt \vrule         \cr
						\noalign{\hrule}
						& Fall-out                               & 0.01467505        & \multispan2\kern3pt 0.01048218\hfill\kern3pt \vrule                & \multispan2\kern3pt 0.00628931\hfill\kern3pt \vrule            & \multispan2\kern3pt 0.03144654\hfill\kern3pt \vrule                & \multispan2\kern3pt 0.07127883\hfill\kern3pt \vrule         \cr
						\noalign{\hrule}
						& Training Time                          & 36.05             & \multispan2\kern3pt 2250.9\hfill\kern3pt \vrule                    & \multispan2\kern3pt 2053.89\hfill\kern3pt \vrule               & \multispan2\kern3pt 38.41\hfill\kern3pt \vrule                     & \multispan2\kern3pt 1362.27\hfill\kern3pt \vrule            \cr
						\noalign{\hrule}
						& \omit\kern3pt (c)\hfill\kern3pt \vrule & Normal- DoS UDP   & \multispan2\kern3pt Normal- OS Fingerprinting\hfill\kern3pt \vrule & \multispan2\kern3pt Normal  Service Scan\hfill\kern3pt \vrule & \multispan2\kern3pt Normal- Data exfiltration\hfill\kern3pt \vrule & \multispan2\kern3pt Normal- keylogging\hfill\kern3pt \vrule \cr
						\noalign{\hrule}
						& Accuracy                               & 0.99998645        & \multispan2\kern3pt 0.99173509\hfill\kern3pt \vrule                & \multispan2\kern3pt 0.99568199\hfill\kern3pt \vrule            & \multispan2\kern3pt 0.98757764\hfill\kern3pt \vrule                & \multispan2\kern3pt 0.98727273\hfill\kern3pt \vrule         \cr
						\noalign{\hrule}
						& Precision                              & 0.99999419        & \multispan2\kern3pt 0.99842856\hfill\kern3pt \vrule                & \multispan2\kern3pt 0.99859012\hfill\kern3pt \vrule            & \multispan2\kern3pt 0\hfill\kern3pt \vrule                         & \multispan2\kern3pt 0.98529412\hfill\kern3pt \vrule         \cr
						\noalign{\hrule}
						& Recall                                 & 0.99999226        & \multispan2\kern3pt 0.99307804\hfill\kern3pt \vrule                & \multispan2\kern3pt 0.99706156\hfill\kern3pt \vrule            & \multispan2\kern3pt 0\hfill\kern3pt \vrule                         & \multispan2\kern3pt 0.91780822\hfill\kern3pt \vrule         \cr
						\noalign{\hrule}
						& Fall-out                               & 0.01257862        & \multispan2\kern3pt 0.05870021\hfill\kern3pt \vrule                & \multispan2\kern3pt 0.21593291\hfill\kern3pt \vrule            & \multispan2\kern3pt 0\hfill\kern3pt \vrule                         & \multispan2\kern3pt 0.00209644\hfill\kern3pt \vrule         \cr
						\noalign{\hrule}
						& Training Time                          & 2269.61           & \multispan2\kern3pt 74.21\hfill\kern3pt \vrule                     & \multispan2\kern3pt 188.27\hfill\kern3pt \vrule                & \multispan2\kern3pt 38.41\hfill\kern3pt \vrule                     & \multispan2\kern3pt 38.74\hfill\kern3pt \vrule              \cr
						\noalign{\hrule}
					}
				}$$
				
			\end{table}
			\end{minipage}
	\end{turn}	
				\begin{center}
					\begin{small}
						\begin{table}[H]	
							\caption{\normalsize Confusion matrices of RNN models for subcategory evaluation.}
							\label{table:RNN_Confusion_Matrices_Sub_Cat_Eval}

							\begin{minipage}{.5\linewidth}
								\addtocounter{table}{-1}
								\begin{longtable}{p{.3\textwidth}|p{.28\textwidth}|l}
									True\textbackslash{}Predict & Normal (0) & \multicolumn{1}{p{.2\textwidth}}{DDoS TCP (1)} \\ \hline
									Normal (0) & 472 & 5  \\ \hline
									DDoS TCP (1) & 74 & 977306                        
									
								\end{longtable}
							\end{minipage}
							\begin{minipage}{.5\linewidth}		
								\addtocounter{table}{-1}			
								\begin{longtable}{||p{.3\textwidth}|p{.28\textwidth}|l}
									True\textbackslash{}Predict & Normal (0) & \multicolumn{1}{p{.2\textwidth}}{DDoS HTTP (1)} \\ \hline
									Normal (0) & 470 & 7  \\ \hline
									DDoS HTTP (1) & 3 & 986                        
									
									\end{longtable}
							\end{minipage}
							\begin{minipage}{.5\linewidth}
								\addtocounter{table}{-1}
								\begin{longtable}{p{.3\textwidth}|p{.28\textwidth}|l}
									True\textbackslash{}Predict & Normal (0) & \multicolumn{1}{p{.2\textwidth}}{DDoS UDP (1)} \\ \hline
									Normal (0) & 474 & 3  \\ \hline
									DDoS UDP (1) & 2 & 948253                        
									
								\end{longtable}
							\end{minipage}
							\begin{minipage}{.5\linewidth}	
								\addtocounter{table}{-1}				
								\begin{longtable}{||p{.3\textwidth}|p{.28\textwidth}|l}
									True\textbackslash{}Predict & Normal (0) & \multicolumn{1}{p{.2\textwidth}}{DoS TCP (1)} \\ \hline
									Normal (0) & 443 & 34  \\ \hline
									DoS TCP (1) & 5 & 615795                        
									
								\end{longtable}
							\end{minipage}
							\begin{minipage}{.5\linewidth}
								\addtocounter{table}{-1}
								\begin{longtable}{p{.3\textwidth}|p{.28\textwidth}|l}
									True\textbackslash{}Predict & Normal (0) & \multicolumn{1}{p{.28\textwidth}}{DoS HTTP (1)} \\ \hline
									Normal (0) & 462 & 15  \\ \hline
									DoS HTTP (1) & 23 & 1462                        
									
								\end{longtable}
							\end{minipage}
							\begin{minipage}{.5\linewidth}	
								\addtocounter{table}{-1}				
								\begin{longtable}{||p{.3\textwidth}|p{.28\textwidth}|l}
									True\textbackslash{}Predict & Normal (0) & \multicolumn{1}{p{.2\textwidth}}{DoS UDP (1)} \\ \hline
									Normal (0) & 471 & 6  \\ \hline
									DoS UDP (1) & 8 & 1032967                        
									
								\end{longtable}
							\end{minipage}
							\begin{minipage}{.5\linewidth}
								\addtocounter{table}{-1}
								\begin{longtable}{p{.3\textwidth}|p{.28\textwidth}|l}
									True\textbackslash{}Predict & Normal (0) & \multicolumn{1}{p{.23\textwidth}}{OS Fingr/t (1)} \\ \hline
									Normal (0) & 449 & 28  \\ \hline
									OS Fingr/t (1) & 124 & 17790                        
									
								\end{longtable}
							\end{minipage}
							\begin{minipage}{.5\linewidth}	
								\addtocounter{table}{-1}				
								\begin{longtable}{||p{.3\textwidth}|p{.28\textwidth}|l}
									True\textbackslash{}Predict & Normal (0) & \multicolumn{1}{p{.2\textwidth}}{Service Scan (1)} \\ \hline
									Normal (0) & 374 & 103  \\ \hline
									Service Scan (1) & 215 & 72953                        
									
								\end{longtable}
							\end{minipage}
							\begin{minipage}{.5\linewidth}
								\addtocounter{table}{-1}
								\begin{longtable}{p{.3\textwidth}|p{.28\textwidth}|l}
									True\textbackslash{}Predict & Normal (0) & \multicolumn{1}{p{.23\textwidth}}{Data ex-filtration (1)} \\ \hline
									Normal (0) & 477 & 0  \\ \hline
									Data exfilt. (1) & 6 & 0                        
									
								\end{longtable}
							\end{minipage}
							\begin{minipage}{.5\linewidth}		
								\addtocounter{table}{-1}			
								\begin{longtable}{||p{.3\textwidth}|p{.28\textwidth}|l}
									True\textbackslash{}Predict & Normal (0) & \multicolumn{1}{p{.2\textwidth}}{Key-logging (1)} \\ \hline
									Normal (0) & 476 & 1  \\ \hline
									Keylogging (1) & 6 & 67                        
									
								\end{longtable}
							\end{minipage}
							
						\end{table}
					\end{small}
				\end{center}
				
			\subsubsection{Overview of three classifiers and discussion of results}
			Overall, results indicate high accuracy both in Binary and Multiclass classification. Data exfiltration had the worst metrics of all (in multiclass classification). Training time is somewhat proportional to the records used to train the model, more records, more time for training, the one that took longest to train was DoS UDP-Normal traffic.
			\begin{center}				
				\begin{small}
					
					\begin{longtable}{|l|p{0.16\textwidth}|p{0.08\textwidth}|p{0.07\textwidth}|p{0.07\textwidth}|l|p{0.16\textwidth}|p{0.08\textwidth}|}
						\caption{\normalsize Summarization of models’ parameters}
						\label{table:SummarizationOfModelsParameters}\\
						\cline{3-8}
						\multicolumn{2}{l|}{}                                                                                                     & Max Iterations & Epochs & Layers & Neurons                                                                                                                                                                                                                                                                      & Activation function                                                                                     & Batch size \\ \hline
						\multicolumn{1}{|l|}{\multirow{2}{*}{SVM}}  & \begin{tabular}[c]{@{}l@{}}10-Best\\   Features\end{tabular}               & 3000           & -      & -      & -                                                                                                                                                                                                                                                                            & -                                                                                                       & -          \\ \cline{2-8} 
						\multicolumn{1}{|l|}{}                      & \begin{tabular}[c]{@{}l@{}}Full\\   Features\end{tabular}                  & 1000           & -      & -      & -                                                                                                                                                                                                                                                                            & -                                                                                                       & -          \\ \hline
						\multicolumn{1}{|l|}{\multirow{3}{*}{RNN}}  & \begin{tabular}[c]{@{}l@{}}10-Best\\   Features\end{tabular}               & -              & 4      & 6      & \begin{tabular}[c]{@{}l@{}}10 Input\\   hidden layers \\  20 1st\\   60 2nd\\  80 3rd\\   90 4th\\    1 Output\\  \end{tabular}                                                                                     & \begin{tabular}[c]{@{}l@{}}Hidden layers:\\ 'tanh'\\ \\   Output layer:\\ 'sigmoid'\end{tabular}    & 100        \\ \cline{2-8} 
						\multicolumn{1}{|l|}{}                      & \begin{tabular}[c]{@{}l@{}}Full\\   Features\end{tabular}                  & -              & 4      & 6      & \begin{tabular}[c]{@{}l@{}}35 Input\\  hidden layers\\   20 1st\\   60 2nd\\  80 3rd\\  90 4th\\   1 Output\\  \end{tabular}                                                                                     & \begin{tabular}[c]{@{}l@{}}Hidden layers:\\ 'tanh'\\ \\   Output layer:\\  'sigmoid'\end{tabular}    & 100        \\ \cline{2-8} 
						\multicolumn{1}{|l|}{}                      & \begin{tabular}[c]{@{}l@{}}Attacks\\ Binary\\ Classifications\end{tabular} & -              & 4      & 9      & \begin{tabular}[c]{@{}l@{}}10 Input\\ hidden layers\\  30 1st\\   40 2nd\\ 40 3rd\\   60 4th\\   80 5th\\   80 6th\\    90 7th\\   1 Output\\ \end{tabular} & \begin{tabular}[c]{@{}l@{}}Hidden layers:\\ 'relu'\\ \\   Output layer:\\ 'sigmoid'\end{tabular} & $\sim$100  \\ \hline
						\multicolumn{1}{|l|}{\multirow{2}{*}{LSTM}} & \begin{tabular}[c]{@{}l@{}}10-Best\\   Features\end{tabular}               & -              & 4      & 6      & \begin{tabular}[c]{@{}l@{}}10 Input\\    hidden layers\\   20 1st\\   60 2nd\\   80 3rd\\  90 4th\\   1 Output\\ \end{tabular}                                                                                     & \begin{tabular}[c]{@{}l@{}}Hidden layers:\\ 'tanh'\\ \\  Output layer:\\ 'sigmoid'\end{tabular}  & 100        \\ \cline{2-8} 
						\multicolumn{1}{|l|}{}                      & \begin{tabular}[c]{@{}l@{}}Full\\   Features\end{tabular}                  & -              & 4      & 6      & \begin{tabular}[c]{@{}l@{}}35 Input\\  hidden layers\\  20 1st\\    60 2nd\\  80 3rd\\     90 4th\\   1 Output\\   \end{tabular}                                                                                     & \begin{tabular}[c]{@{}l@{}}Hidden layers:\\ 'tanh'\\ \\  Output layer: \\'sigmoid'\end{tabular}  & 100        \\ \hline
					\end{longtable}
				\end{small}
			\end{center}
			Additionally, in the binary classification, Fall-out values were rather high, with the exception of RNN and LSTM models for the fully-featured version of the dataset. This could be explained by a number of factors, such as poor optimization of models in use and the relatively low number of epochs that we chose, in order to speed up the process. In table \ref{table:SummarizationOfModelsParameters}, the parameters that were chosen for the three models are presented.
			
	\section{Conclusion}
	This paper presents a new dataset, Bot-IoT, which incorporates both normal IoT-related and other network traffic, along with various types of attack traffic commonly used by botnets. This dataset was developed on a realistic testbed, and has been labeled, with the label features indicated an attack flow, the attacks category and subcategory for possible multiclass classification purposes. Additional features were generated to enhance the predictive capabilities of classifiers trained on this model. Through statistical analysis, a subset of the original dataset was produced, comprised of the 10-best features. Finally, four metrics were used in order to compare the validity of the dataset, specifically Accuracy, Precision, Recall, Fall-out. We observed the highest accuracy and recall from the SVM model that was trained on the full-featured dataset, while the highest precision and lowest fall-out from the SVM model of the of the 10-best feature dataset version. With further optimization of these models, we argue that better results could be achieved. In Future, we plan to develop a network forensic model using deep learning and evaluate its reliability using the BoT-IoT dataset. 

	\bibliography{Bot-IoT_Dataset}

\begin{thebibliography}{10}
\expandafter\ifx\csname url\endcsname\relax
  \def\url#1{\texttt{#1}}\fi
\expandafter\ifx\csname urlprefix\endcsname\relax\def\urlprefix{URL}\fi
\expandafter\ifx\csname href\endcsname\relax
  \def\href#1#2{#2} \def\path#1{#1}\fi

\bibitem{2018q}
\href{https://www.unsw.adfa.edu.au/unsw-canberra-cyber/cybersecurity/ADFA-NB15-Datasets/bot_iot.php}{Bot-iot}
  (2018).
\newline\urlprefix\url{https://www.unsw.adfa.edu.au/unsw-canberra-cyber/cybersecurity/ADFA-NB15-Datasets/bot_iot.php}

\bibitem{ronen2017iot}
E.~Ronen, A.~Shamir, A.-O. Weingarten, C.~O’Flynn, Iot goes nuclear: Creating
  a zigbee chain reaction, in: Security and Privacy (SP), 2017 IEEE Symposium
  on, IEEE, 2017, pp. 195--212.

\bibitem{kolias2017ddos}
C.~Kolias, G.~Kambourakis, A.~Stavrou, J.~Voas, Ddos in the iot: Mirai and
  other botnets, Computer 50~(7) (2017) 80--84.

\bibitem{pimenta2017cybersecurity}
G.~A. Pimenta~Rodrigues, R.~de~Oliveira~Albuquerque, F.~E. Gomes~de Deus,
  et~al., Cybersecurity and network forensics: Analysis of malicious traffic
  towards a honeynet with deep packet inspection, Applied Sciences 7~(10)
  (2017) 1082.

\bibitem{liu2015external}
C.~Liu, C.~Yang, X.~Zhang, J.~Chen, External integrity verification for
  outsourced big data in cloud and iot: A big picture, Future generation
  computer systems 49 (2015) 58--67.

\bibitem{grajeda2017availability}
C.~Grajeda, F.~Breitinger, I.~Baggili, Availability of datasets for digital
  forensics--and what is missing, Digital Investigation 22 (2017) S94--S105.

\bibitem{2018}
\href{http://kdd.ics.uci.edu/databases/kddcup99/kddcup99.html}{Kddcup99
  dataset}.
\newline\urlprefix\url{http://kdd.ics.uci.edu/databases/kddcup99/kddcup99.html}

\bibitem{sharafaldin2018toward}
I.~Sharafaldin, A.~H. Lashkari, A.~A. Ghorbani, Toward generating a new
  intrusion detection dataset and intrusion traffic characterization, in:
  Proceedings of fourth international conference on information systems
  security and privacy, ICISSP, 2018.

\bibitem{moustafa2015unsw}
N.~Moustafa, J.~Slay, Unsw-nb15: a comprehensive data set for network intrusion
  detection systems (unsw-nb15 network data set), in: Military Communications
  and Information Systems Conference (MilCIS), 2015, IEEE, 2015, pp. 1--6.

\bibitem{2018a}
\href{https://www.ll.mit.edu/ideval/data/1998data.html}{1998 darpa intrusion
  detection evaluation data set}.
\newline\urlprefix\url{https://www.ll.mit.edu/ideval/data/1998data.html}

\bibitem{gubbi2013internet}
J.~Gubbi, R.~Buyya, S.~Marusic, M.~Palaniswami, Internet of things (iot): A
  vision, architectural elements, and future directions, Future generation
  computer systems 29~(7) (2013) 1645--1660.

\bibitem{silva2013botnets}
S.~S. Silva, R.~M. Silva, R.~C. Pinto, R.~M. Salles, Botnets: A survey,
  Computer Networks 57~(2) (2013) 378--403.

\bibitem{khattak2014taxonomy}
S.~Khattak, N.~R. Ramay, K.~R. Khan, A.~A. Syed, S.~A. Khayam, A taxonomy of
  botnet behavior, detection, and defense, IEEE communications surveys \&
  tutorials 16~(2) (2014) 898--924.

\bibitem{amini2015survey}
P.~Amini, M.~A. Araghizadeh, R.~Azmi, A survey on botnet: classification,
  detection and defense, in: Electronics Symposium (IES), 2015 International,
  IEEE, 2015, pp. 233--238.

\bibitem{palmer2001road}
G.~Palmer, A road map for digital forensic research: Report from the first
  digital forensic workshop, 7--8 august 2001, DFRWS Technical Report
  DTR-T001-01.

\bibitem{hodo2016threat}
E.~Hodo, X.~Bellekens, A.~Hamilton, P.-L. Dubouilh, E.~Iorkyase, C.~Tachtatzis,
  R.~Atkinson, Threat analysis of iot networks using artificial neural network
  intrusion detection system, in: Networks, Computers and Communications
  (ISNCC), 2016 International Symposium on, IEEE, 2016, pp. 1--6.

\bibitem{garcia2009anomaly}
P.~Garcia-Teodoro, J.~Diaz-Verdejo, G.~Maci{\'a}-Fern{\'a}ndez, E.~V{\'a}zquez,
  Anomaly-based network intrusion detection: Techniques, systems and
  challenges, computers \& security 28~(1-2) (2009) 18--28.

\bibitem{wang2016attack}
K.~Wang, M.~Du, Y.~Sun, A.~Vinel, Y.~Zhang, Attack detection and distributed
  forensics in machine-to-machine networks, IEEE Network 30~(6) (2016) 49--55.

\bibitem{rieck2008learning}
K.~Rieck, T.~Holz, C.~Willems, P.~D{\"u}ssel, P.~Laskov, Learning and
  classification of malware behavior, in: International Conference on Detection
  of Intrusions and Malware, and Vulnerability Assessment, Springer, 2008, pp.
  108--125.

\bibitem{doshi2018machine}
R.~Doshi, N.~Apthorpe, N.~Feamster, Machine learning ddos detection for
  consumer internet of things devices, arXiv preprint arXiv:1804.04159.

\bibitem{nguyen2008survey}
T.~T. Nguyen, G.~Armitage, A survey of techniques for internet traffic
  classification using machine learning, IEEE Communications Surveys \&
  Tutorials 10~(4) (2008) 56--76.

\bibitem{de2001mining}
O.~De~Vel, A.~Anderson, M.~Corney, G.~Mohay, Mining e-mail content for author
  identification forensics, ACM Sigmod Record 30~(4) (2001) 55--64.

\bibitem{alomari2014design}
E.~Alomari, S.~Manickam, B.~Gupta, P.~Singh, M.~Anbar, Design, deployment and
  use of http-based botnet (hbb) testbed, in: Advanced Communication Technology
  (ICACT), 2014 16th International Conference on, IEEE, 2014, pp. 1265--1269.

\bibitem{carl2006using}
L.~Carl, et~al., Using machine learning technliques to identify botnet traffic,
  in: Local Computer Networks, Proceedings 2006 31st IEEE Conference on. IEEE,
  2006.

\bibitem{bhatia2014framework}
S.~Bhatia, D.~Schmidt, G.~Mohay, A.~Tickle, A framework for generating
  realistic traffic for distributed denial-of-service attacks and flash events,
  Computers \& Security 40 (2014) 95--107.

\bibitem{behal2017detection}
S.~Behal, K.~Kumar, Detection of ddos attacks and flash events using
  information theory metrics--an empirical investigation, Computer
  Communications 103 (2017) 18--28.

\bibitem{emaN}
\href{https://ostinato.org/}{Ostinato}.
\newline\urlprefix\url{https://ostinato.org/}

\bibitem{soni2017survey}
D.~Soni, A.~Makwana, A survey on mqtt: a protocol of internet of things (iot),
  in: Proceeding of the International Conference on Telecommunication, Power
  Analysis and Computing Techniques, Chennai: IN, 2017.

\bibitem{brugger2007assessment}
S.~T. Brugger, J.~Chow, An assessment of the darpa ids evaluation dataset using
  snort, UCDAVIS department of Computer Science 1~(2007) (2007) 22.

\bibitem{fernandezugr}
G.~M. Fern{\'a}ndez, J.~Camacho, R.~Mag{\'a}n-Carri{\'o}n,
  P.~Garc{\i}a-Teodoro, R.~Theron, Ugr’16: A new dataset for the evaluation
  of cyclostationarity-based network idss.

\bibitem{tavallaee2009detailed}
M.~Tavallaee, E.~Bagheri, W.~Lu, A.~A. Ghorbani, A detailed analysis of the kdd
  cup 99 data set, in: Computational Intelligence for Security and Defense
  Applications, 2009. CISDA 2009. IEEE Symposium on, IEEE, 2009, pp. 1--6.

\bibitem{UNIBS2009}
\href{http://www.ing.unibs.it/ntw/tools/traces/}{Unibs, university of brescia
  dataset} (2009).
\newline\urlprefix\url{http://www.ing.unibs.it/ntw/tools/traces/}

\bibitem{bhuyan2015towards}
M.~H. Bhuyan, D.~K. Bhattacharyya, J.~K. Kalita, Towards generating real-life
  datasets for network intrusion detection., IJ Network Security 17~(6) (2015)
  683--701.

\bibitem{2018e}
\href{https://www.caida.org/data/}{Center of applied internet data analysis}.
\newline\urlprefix\url{https://www.caida.org/data/}

\bibitem{2005}
\href{http://www.icir.org/enterprise-tracing/}{Lawrence berkley national
  laboratory (lbnl), icsi, lbnl/icsi enterprise tracing project} (2005).
\newline\urlprefix\url{http://www.icir.org/enterprise-tracing/}

\bibitem{CanadianInstituteofCybersecurity2018}
U.~o. n.~B. Canadian Institute~of Cybersecurity,
  \href{http://www.unb.ca/cic/datasets/index.html}{Iscx dataset}.
\newline\urlprefix\url{http://www.unb.ca/cic/datasets/index.html}

\bibitem{ammar2015decision}
A.~Ammar, A decision tree classifier for intrusion detection priority tagging,
  Journal of Computer and Communications 3~(04) (2015) 52.

\bibitem{gogoi2012packet}
P.~Gogoi, M.~H. Bhuyan, D.~Bhattacharyya, J.~K. Kalita, Packet and flow based
  network intrusion dataset, in: International Conference on Contemporary
  Computing, Springer, 2012, pp. 322--334.

\bibitem{2018i}
\href{https://nodered.org/}{Node-red tool}.
\newline\urlprefix\url{https://nodered.org/}

\bibitem{2018j}
\href{https://qosient.com/argus/index.shtml}{Argus tool}.
\newline\urlprefix\url{https://qosient.com/argus/index.shtml}

\bibitem{2018k}
\href{https://www.vmware.com/au/products/esxi-and-esx.html}{Esxi hypervisor}.
\newline\urlprefix\url{https://www.vmware.com/au/products/esxi-and-esx.html}

\bibitem{2018l}
\href{https://www.vmware.com/au/products/vsphere.html}{vsphere client}.
\newline\urlprefix\url{https://www.vmware.com/au/products/vsphere.html}

\bibitem{2018m}
\href{https://aws.amazon.com/iot-core/features/}{Iot hub aws}.
\newline\urlprefix\url{https://aws.amazon.com/iot-core/features/}

\bibitem{2018n}
\href{https://mosquitto.org/}{Mosquitto mqtt broker}.
\newline\urlprefix\url{https://mosquitto.org/}

\bibitem{hall2015pearson}
G.~Hall, Pearson’s correlation coefficient, other words 1~(9).

\bibitem{lesne2014shannon}
A.~Lesne, Shannon entropy: a rigorous notion at the crossroads between
  probability, information theory, dynamical systems and statistical physics,
  Mathematical Structures in Computer Science 24~(3).

\bibitem{2018o}
\href{https://packages.ubuntu.com/search?keywords=cron}{Cron scheduling
  package}.
\newline\urlprefix\url{https://packages.ubuntu.com/search?keywords=cron}

\bibitem{2018p}
\href{https://www.wireshark.org/}{Tshark network analysis tool}.
\newline\urlprefix\url{https://www.wireshark.org/}

\bibitem{2018b}
\href{https://qosient.com/argus/}{Argus (audit record generation and
  utilization system)}.
\newline\urlprefix\url{https://qosient.com/argus/}

\bibitem{paliwal2012denial}
S.~Paliwal, R.~Gupta, Denial-of-service, probing \& remote to user (r2l) attack
  detection using genetic algorithm, International Journal of Computer
  Applications 60~(19) (2012) 57--62.

\bibitem{bartlett2007understanding}
G.~Bartlett, J.~Heidemann, C.~Papadopoulos, Understanding passive and active
  service discovery (extended), Tech. rep., Technical Report ISI-TR-2007-642,
  USC/Information Sciences Institute (2007).

\bibitem{hoque2014network}
N.~Hoque, M.~H. Bhuyan, R.~C. Baishya, D.~K. Bhattacharyya, J.~K. Kalita,
  Network attacks: Taxonomy, tools and systems, Journal of Network and Computer
  Applications 40 (2014) 307--324.

\bibitem{lyon2009nmap}
G.~F. Lyon, Nmap network scanning: The official Nmap project guide to network
  discovery and security scanning, Insecure, 2009.

\bibitem{2018c}
\href{http://www.hping.org}{hping}.
\newline\urlprefix\url{http://www.hping.org}

\bibitem{2018d}
\href{https://www.aldeid.com/wiki/Xprobe2}{Xprobe2}.
\newline\urlprefix\url{https://www.aldeid.com/wiki/Xprobe2}

\bibitem{zargar2013survey}
S.~T. Zargar, J.~Joshi, D.~Tipper, A survey of defense mechanisms against
  distributed denial of service (ddos) flooding attacks, IEEE communications
  surveys \& tutorials 15~(4) (2013) 2046--2069.

\bibitem{tankard2011advanced}
C.~Tankard, Advanced persistent threats and how to monitor and deter them,
  Network security 2011~(8) (2011) 16--19.

\bibitem{jesudoss2014survey}
A.~Jesudoss, N.~Subramaniam, A survey on authentication attacks and
  countermeasures in a distributed environment, Indian J Comput Sci Eng IJCSE 5
  (2014) 71--77.

\bibitem{2018h}
\href{https://www.metasploit.com/}{Metasploit framework}.
\newline\urlprefix\url{https://www.metasploit.com/}

\bibitem{2018f}
\href{http://manpages.ubuntu.com/manpages/xenial/man8/logkeys.8.html}{Logkeys
  software}.
\newline\urlprefix\url{http://manpages.ubuntu.com/manpages/xenial/man8/logkeys.8.html}

\bibitem{2018g}
\href{https://packages.ubuntu.com/trusty/net/hydra}{Hydra software}.
\newline\urlprefix\url{https://packages.ubuntu.com/trusty/net/hydra}

\bibitem{zheng2011feature}
Y.~Zheng, C.~K. Kwoh, A feature subset selection method based on
  high-dimensional mutual information, Entropy 13~(4) (2011) 860--901.

\bibitem{meyer2001support}
D.~Meyer, F.~T. Wien, Support vector machines, R News 1~(3) (2001) 23--26.

\bibitem{grossberg2013recurrent}
S.~Grossberg, Recurrent neural networks, Scholarpedia 8~(2) (2013) 1888.

\bibitem{greff2017lstm}
K.~Greff, R.~K. Srivastava, J.~Koutn{\'\i}k, B.~R. Steunebrink, J.~Schmidhuber,
  Lstm: A search space odyssey, IEEE transactions on neural networks and
  learning systems 28~(10) (2017) 2222--2232.

\end{thebibliography}
\end{document}